\documentclass[11pt]{article}
\usepackage{amsfonts}
\usepackage[english]{babel}
\usepackage{graphicx}
\usepackage{epstopdf}
\usepackage{epsfig}
\usepackage{amssymb}
\usepackage{setspace}
\usepackage{caption}
\usepackage{amsfonts}
\usepackage{color}
\usepackage{amsmath}
\usepackage{float}
\usepackage{comment}
\newcommand {\be}{\begin{equation}}
\newcommand {\ee}{\end{equation}}
\newcommand {\bea}{\begin{array}}
	
	\newcommand {\eea}{\end{array}}
\newcommand{\sch}{Schwarzschild~}

\begin{document}

\begin{titlepage}
	\vspace{1cm}
	\begin{center}
		{\Large \bf {Magnetized Taub-NUT spacetime}}\\
	\end{center}
	\vspace{2cm}
	\begin{center}
		\renewcommand{\thefootnote}{\fnsymbol{footnote}}
		Haryanto M. Siahaan{\footnote{haryanto.siahaan@unpar.ac.id}}\\
		Center for Theoretical Physics,\\
		Department of Physics, Parahyangan Catholic University,\\
		Jalan Ciumbuleuit 94, Bandung 40141, Indonesia
		\renewcommand{\thefootnote}{\arabic{footnote}}
	\end{center}
	
	\begin{abstract}
		
	We find an exact solution describing the spacetime outside a massive object with NUT parameter embedded in an external magnetic field. To get the solution, we employ the Ernst magnetization transformation to the Taub-NUT metric as the seed solution. The massless limit of this new solution is the Melvin-Taub-NUT spacetime. Some aspects in the magnetized Taub-NUT spacetime are investigated, such as the surface geometry, the existence of closed time like curve, and the electromagnetic properties.
		
	\end{abstract}
\end{titlepage}\onecolumn
\bigskip

\section{Introduction}
\label{sec:intro}

Exact solutions in the Einstein-Maxwell theory are always fascinating objects to be studied \cite{Stephani:2003tm,IslamBook,Wald:1984rg.book,Misner:1974qy}. It starts from the mathematical aspects of the solution to some possible astrophysical-related phenomenon. In the Einstein-Maxwell theory the spacetime can contain electromagnetic fields, and sometime is known as the electrovacuum system. The most general asymptotically flat spacetime solution in Einstein-Maxwell theory that contains a black hole is known as the Kerr-Newman solution. This solution describes a black hole with rotation and electric charge as well. Despite it is very unlikely for a collapsing object to maintain a significant amount of electric charge, this type of solution has been discussed significantly in literature \cite{Stephani:2003tm,Wald:1984rg.book,Griffiths:2009dfa}.

Another interesting solution of black hole spacetime containing electromagnetic fields is the magnetized solution proposed by Wald \cite{Wald:1974np}. The solution by Wald describes a black hole immersed in a uniform magnetic field, where the Maxwell field solution is generated by the associated Killing symmetries of spacetime. However, in Wald's prescription, the Maxwell field is considered as some perturbations in the spacetime. The nonperturbative consideration of a black hole immersed in a homogeneous magnetic field was introduced by Ernst \cite{Ernst}. Ersnt solution for magnetized black hole can be viewed as an embedding a black hole in Melvin magnetic universe where the spacetime outside black hole is filled by an homogeneous magnetic field. Ernst method uses a Harison type of transformation \cite{Harrison} that is applied to a known seed solution in Einstein-Maxwell theory. Nevertheless, the resulting magnetized spacetimes are no longer flat at asymptotic despite resulting from an asymptotically flat seed metric. Various aspects of black holes immersed in external magnetic field had been studied extensively in literature \cite{Aliev:1989wx,Kolos:2015iva,Tursunov:2014loa,Aliev:1989wz,Aliev:1980et,Aliev:1986wu,Aliev:1988wy,Aliev:1989sw,Galtsov:1978ag,Hiscock:1980zf,Orekhov:2016bpc,Gibbons:2013yq,Gibbons:2013dna,Rogatko:2016knj}.

Despite the loss of asymptotic flatness for a magnetized spacetime containing a black hole, this solution has been considered to have some astrophysical uses especially in modeling the spacetime near rotating supermassive black hole surrounded by hot moving plasma \cite{Aliev:1989wx}. Indeed, a full comprehension for the interaction of a black hole and its surrounding magnetic field due to the accretion disc requires a sophisticated general relativistic treatment, at least by employing a costly numerical approach. If the full general relativistic or the comprehensive numerical treatment is not necessary, in the sense that we just seek for some approximate qualitative explanations, the picture of magnetized black hole by Wald or even the non-perturbative model by Ernst can be the alternatives. For example, these models can explain the charge induction by a black hole immersed in the magnetic field, or the Meissner-like effect that may occur for this type of black hole \cite{Bicak:2015lxa}. In particular, the superradiant instability of a magnetized rotating black hole is studied in \cite{Brito:2014nja}. Another aspects had been investigated as well, such as the Kerr/CFT correspondence for some magnetized black holes in \cite{Siahaan:2015xia,Astorino:2015lca,Astorino:2015naa}, and some conserved quantities calculations are proposed in \cite{Astorino:2016hls}.

In the vacuum Einstein system, the Taub-NUT solution generalizes \sch solution to include the so-called NUT (Newman-Unti-Tamburino) parameter $l$. This parameter is interpreted as a ``gravitomagnetic'' mass in an analogy to the magnetic monopole in electromagnetic theory. However, the presence of NUT parameter yields several new features in spacetime compared to the null NUT parameter counterpart \cite{Griffiths:2009dfa}. First is the loss of asymptotic flatness, for the asymptotically flat null NUT case. Indeed, this asymptotic behavior demands some delicate approaches in defining the conserved charges in the spacetime. Second, the non-existence of physical singularity at the origin. This is interesting since it leads to the question of defining black hole in such spacetime, which normally we understand as a true singularity covered by a horizon. Instead of a true singularity at origin, the spacetime with NUT parameter possesses the conic singularity on its symmetry axis which gives rise to a problem in describing the black hole horizon. Despite these issues regarding black hole picture in a spacetime with NUT parameter, discussions on Kerr-Newman-Taub-NUT black hole family are still among the active areas in gravitational studies \cite{Chakraborty:2017nfu,Pradhan:2014zia,Sakti:2017pmt,Sakti:2019krw,Sakti:2017pmt,Aliev:2008wv,Duztas:2017lxk,Cebeci:2015fie,Zakria:2015eua,Mukherjee:2018dmm,Abdujabbarov:2011uv,Abdujabbarov:2012jj,Ahmedov:2011wb,Abdujabbarov:2008mz,Abdujabbarov:2012bn}. In particular, the discussion related to the circular motion in a spacetime with NUT parameter \cite{Jefremov:2016dpi} where the authors found that the existence of NUT parameter leads to a new non trivial constraint to the equatorial circular motion for a test body. This problem also occurs in gravitational theories beyond Einstein-Maxwell, for example in low energy string theory \cite{Siahaan:2019kbw} and braneworld scenario \cite{Siahaan:2020bga}. In a recent work \cite{Ciambelli:2020qny}, the authors show that the Misner string contribution to the entropy of Taub-NUT-AdS can be renormalized by introducing the Gauss-Bonnet term, and in \cite{Cisterna:2021xxq} the authors show how to embed the Taub-NUT solutions in general scalar-tensor theories.

There exist magnetization of some well-known solution in Einstein-Maxwell theory whose aspects have been studied extensively \cite{Griffiths:2009dfa,Aliev:1989wx}. In this work, we introduce a new solution namely the magnetized Taub-NUT spacetime. The idea is straightforward, i.e. performing the magnetization transformation to the Taub-NUT metric as the seed solution. One key aspect is the compatibility of the Taub-NUT metric with the Lewis-Papapetrou-Weyl (LPW) line element. The obtained solution describes an object with mass and NUT parameter embedded in an external magnetic field. A quite similar idea where the weak external magnetic field exists outside an object with NUT parameter has been performed in \cite{Frolov:2017bdq}.

The properties of the event horizon under the influence of some external magnetic fields are also an interesting aspect to be investigated \cite{Wild:1980zz}. It is known for the magnetized \sch solution, the scalar curvature of the horizon varies depending on the strength of the external magnetic field. It can take positive, zero, or negative values, and these values associate to some different physics. Recall that a ``normal'' horizon such as the one of \sch black hole has a positive curvature, which is understood from its spherical form. However, despite the shape of the horizon changing due to the presence of an external magnetic field, the total area of the horizon does not vary.

The organization of this paper is as follows. In the next section, we provide some reviews of the Ernst magnetization procedure by using a complex differential operator. In section \ref{sec.3.MagTaubNUT}, after employing the magnetization procedure to the Taub-NUT spacetime, we obtain the magnetized Taub-NUT solution. The surface geometry and closed timelike curve in this new spacetime are discussed in section \ref{sec.4.SurfaceGeom}. In section \ref{sec.EM}, some of the electromagnetic properties in the spacetime are studied. Finally, we give some conclusions and discussions. We consider the natural units $c={\hbar} = k_B = G_4 = 1$.

\section{Magnetization of a spacetime}\label{sec.2.MagnetizationSpacetime}

Magnetization of a spacetime can be done by employing the Ernst magnetization to the metric whose line element can be written as the Lewis-Papapetrou-Weyl (LPW) type,
\be\label{LPWmetric1} 
ds^2  = - f^{ - 1} \left( {\rho ^2 dt^2  - e^{2\gamma } d\zeta d\zeta ^* } \right) + f\left( {d\phi  - \omega dt} \right)^2 \,.
\ee
Above, the $f$, $\gamma$, and $\omega$ are functions of a complex coordinate $\zeta$. Above we are using the $-+++$ signs convention for the spacetime, and the $^*$ notation represents the complex conjugation. In Einstein-Maxwell theory, the metric (\ref{LPWmetric1}) together with a vector solution ${\bf A} = A_t dt + A_\phi d\phi$ obey the field equations
\be \label{eq.Einstein-Maxwell}
{R_{\mu \nu }} = 2{F_{\mu \alpha }}F_\nu ^\alpha  - \frac{1}{2}{g_{\mu \nu }}{F_{\alpha \beta }}{F^{\alpha \beta }}\,,
\ee 
where $R_{\mu\nu}$ is Ricci tensor, and ${F_{\mu \nu }} = {\partial _\mu }{A_\nu } - {\partial _\nu }{A_\mu }$ is the Maxwell field-strength tensor. 

Interestingly, Ernst \cite{ErnstI:1967wx,ErnstII:1967by} showed that the last equations can give us a set of wave-like equations to the Ernst gravitational and electromagnetic potentials. Using the metric functions in (\ref{LPWmetric1}), we can construct the gravitational Ernst potential
\be \label{Ernst.potential.Grav}
{\cal E} = f + \left| {\Phi } \right|^2   - i\Psi \,,
\ee 
where the electromagnetic Ernst potential $\Phi$ consists of
\be \label{Ernst.potential.EM}
\Phi  = A_\phi   + i\tilde A_\phi  \,.
\ee 
Note that the real part of $\Phi$ above is $A_\phi$ instead of $A_t$ as appeared in \cite{ErnstII:1967by} since the gravitational Ernst potential ${\cal E}$ is defined with respect to the Killing vector $\partial_\phi$. The relation between these vector components is given by
\be \label{eqA}
\nabla A_t  =  - \omega \nabla A_\phi   - i\frac{\rho }{f}\nabla \tilde A_\phi  \,,
\ee 
where the twist potential $\Psi$ satisfies
\be \label{eq.Psi}
\nabla \Psi  = \frac{{i f^2 }}{\rho }\nabla \omega + 2i\Phi ^* \nabla \Phi  \,.
\ee 
Interestingly, as it was first shown by Ernst in \cite{ErnstI:1967wx,ErnstII:1967by}, the Einstein-Maxwell eq. (\ref{eq.Einstein-Maxwell}) dictates the Ernst potentials to obey the nonlinear complex equations
\be \label{eq.Ernst.grav}
\left( {{\mathop{\rm Re}\nolimits} \left\{ {\cal E} \right\} + {{\left| \Phi  \right|}^2}} \right)\nabla^2 {\cal E} = \left( {\nabla {\cal E} + 2{\Phi ^*}\nabla \Phi } \right) \cdot \nabla {\cal E}\,,
\ee 
and
\be \label{eq.Ernst.EM}
\left( {{\mathop{\rm Re}\nolimits} \left\{ {\cal E} \right\} + {{\left| \Phi  \right|}^2}} \right)\nabla^2 {\Phi} = \left( {\nabla {\cal E} + 2{\Phi ^*}\nabla \Phi } \right) \cdot \nabla {\Phi}\,.
\ee 

The magnetization procedure according to Ernst can be written as the following,
\be \label{magnetization}
{\cal E} \to {\cal E}' = \Lambda ^{ - 1} {\cal E}~~~{\rm and}~~~\Phi  \to \Phi ' = \Lambda ^{ - 1} \left( {\Phi  - b {\cal E}} \right)\,,
\ee
where 
\be \label{LambdaDEF}
\Lambda  = 1 - 2b\Phi  + b^2 {\cal E}\,.
\ee
In the equations above, the constant $b$ represents the magnetic field strength in the spacetime\footnote{For economical reason, we prefer to express the magnetic parameter as $b$ instead of $B/2$ as appeared in \cite{Ernst}. The relation is $B=2b$.}. The transformation (\ref{magnetization}) leaves the two equations (\ref{eq.Ernst.grav}) and (\ref{eq.Ernst.EM}) remain unchanged. In other words, the new fields $\left\{g'_{\mu\nu},A'_{\mu}\right\}$ after performing the Ernst transformation (\ref{magnetization}) still belong to the field equations (\ref{eq.Einstein-Maxwell}).  

In particular, the transformed line element (\ref{LPWmetric1}) resulting from the magnetization (\ref{magnetization}) has the components
\be \label{fp}
f' = {\mathop{\rm Re}\nolimits} \left\{ {{\cal E}'} \right\} - \left| {\Phi '} \right|^{2}  = \left| \Lambda  \right|^{-2} f\,.
\ee 
The $\omega '$ function obeys
\be \label{wp}
\nabla \omega ' = \left| \Lambda  \right|^2 \nabla \omega  - \frac{\rho }{f}\left( {\Lambda ^* \nabla \Lambda  - \Lambda \nabla \Lambda ^* } \right)\,,
\ee 
and $\gamma$ in (\ref{LPWmetric1}) remains unchanged. Since all the incorporated functions in the metric (\ref{LPWmetric1}) depend only on $\rho$ and $z$ coordinates, then the operator $\nabla$ can be defined in the flat Euclidean space
\be\label{metric2rho.z}
d\zeta d\zeta ^*  = d\rho ^2  + dz^2 \,,
\ee 
where we have set the complex coordinate $d\zeta  = d\rho  + idz$. Here we have $\nabla  = \partial _\rho   + i\partial _z $, accordingly.

As we know, the spacetime solutions in Einstein-Maxwell theory are normally expressed in the Boyer-Lindquist type coordinate $\left\{ {t,r,x = \cos \theta ,\phi } \right\}$. Consequently, the LPW type metric (\ref{LPWmetric1}) with stationary and axial Killing symmetries will have the metric function that depend only on $r$ and $x$ coordinates, and the corresponding flat metric line element reads
\be \label{metric2rx}
d\zeta d\zeta ^*  = \frac{{dr^2 }}{{\Delta _r }} + \frac{{dx^2 }}{{\Delta _x }}\,,
\ee 
where $\Delta _r = \Delta _r \left(r\right)$ and $\Delta _x = \Delta _x \left(x\right)$. Accordingly, the corresponding operator $\nabla$ will read $\nabla  = \sqrt {\Delta _r } \partial _r  + i\sqrt {\Delta _x } \partial _x $. Furthermore, we have $\rho^2 = \Delta_r\Delta_x$ which then allows us to write the components of eq. (\ref{eqA}) as
\be \label{drAt}
\partial _r A_t  =  - \omega \partial _r A_\phi   + \frac{{\Delta _x }}{f}\partial _x \tilde A_\phi  \,,
\ee 
and
\be \label{dxAt}
\partial _x A_t  =  - \omega \partial _x A_\phi   - \frac{{\Delta _r }}{f}\partial _r \tilde A_\phi  \,.
\ee 
The last two equations are useful later in obtaining the $A_t$ component associated to the magnetized spacetime according to (\ref{magnetization}). To complete some details on the magnetization procedure, another equations for the magnetized metric functions are
\be \label{drwp}
\partial _r \omega ' = \left| \Lambda  \right|^2 \partial _r \omega  + \frac{{\Delta _x }}{f}{\mathop{\rm Im}\nolimits} \left\{ {\Lambda ^* \partial _x \Lambda  - \Lambda \partial _x \Lambda ^* } \right\} \,,
\ee 
and
\be \label{dxwp}
\partial _x \omega ' = \left| \Lambda  \right|^2 \partial _x \omega  - \frac{{\Delta _r }}{f}{\mathop{\rm Im}\nolimits} \left\{ {\Lambda ^* \partial _r \Lambda  - \Lambda \partial _r \Lambda ^* } \right\}\,.
\ee 
In the next section, we will employ this magnetization prescription to the Taub-NUT spacetime.

\section{Magnetizing the Taub-NUT spacetime}\label{sec.3.MagTaubNUT}

Taub-NUT spacetime is a non-trivial extension of \sch solution, where in addition to mass parameter $M$, the solution contains an extra parameter $l$ known as the NUT parameter. However, unlike the mass $M$ which can be considered as a conserved quantity due to the timelike Killing symmetry $\partial_t$ \cite{Wald:1984rg.book}, the NUT parameter cannot be viewed in analogous way as a sort of conserved charge associated to a symmetry in the spacetime. The line element of Taub-NUT spacetime can be expressed as \cite{Griffiths:2009dfa}
\be\label{metricTaubNUT}
ds^2  = - \frac{{\Delta _r }}{{r^2  + l^2 }}\left( {dt - 2lxd\phi } \right)^2  + \left( {r^2  + l^2 } \right)\left( {\frac{{dr^2 }}{{\Delta _r }} + \frac{{dx^2 }}{{\Delta _x }}} \right) + \left( {r^2  + l^2 } \right)\Delta _x d\phi ^2 \,,
\ee 
where $\Delta_r = r^2 - 2Mr -l^2$ and $\Delta_x = 1-x^2$. Matching this line element to the LPW form (\ref{LPWmetric1}) gives us
\be \label{f0}
f = \frac{{\Delta _x \left( {r^2  + l^2 } \right)^2  - 4\Delta _r l^2 x^2 }}{{r^2  + l^2 }}\,,
\ee 
\be \label{w0}
\omega  = -\frac{{2\Delta _r lx}}{{4\Delta _r l^2 x^2  - \Delta _x \left( {r^2  + l^2 } \right)^2 }}\,,
\ee 
and
\be \label{e2g0}
e^{2\gamma }  = \Delta _x \left( {r^2  + l^2 } \right)^2  - 4\Delta _r l^2 x^2 \,.
\ee 
Note that from eqs. (\ref{metric2rho.z}) and (\ref{metric2rx}), one can find $\rho ^2  = \Delta_r\Delta_x $ and $z = rx$. Furthermore, by using eq. (\ref{eq.Psi}) one can show that the associated twist potential for Taub-NUT metric can be written as
\be 
\Psi  = -{\frac {2l \left( {r}^{3}+r{l}^{2}+{r}^{3}{x}^{2}-3r{l}^{2}{x}^{2}+M{l}^{2}{x}^{2}-3M{r}^{2}{x}^{2} \right) }{{r}^{2}+{l}^{2}}}\,.
\ee 
Thence, the gravitational Ernst potential (\ref{eq.Ernst.grav}) defined with respect to $\partial_\phi$ for Taub-NUT spacetime (\ref{metricTaubNUT}) is 
\be \label{seedE}
{\cal E} = {\frac {6Mrl{x}^{2}-{r}^{2}l+{l}^{3}-3l{r}^{2}{x}^{2}+3{l}^{3}{x}^{2} + i\left(\Delta_x \left\{r^3 + 3rl^2\right\} +2M{l}^{2}{x}^{2}\right) }{l+ir}}\,,
\ee 
and the electromagnetic Ernst potential $\Phi$ vanishes. Also, from (\ref{LambdaDEF}) we can have 
\be \label{Lambda}
\Lambda = \frac{ \delta_x b^2 {l}^{3}+ \left(1-{b}^{2}{r}^{2}\delta_x+6{b}^{2}Mr{x}^{2} \right) l + i\left\{\left( 3{b}^{2}r\Delta_x+2{b}^{2}M{x}^{2} \right) {l}^{2}+r+{b}^{2}{r}^{3}\Delta_x\right\} }{l+ir}
\ee 
where $\delta_x = 1+3x^2$. Recall that this $\Lambda$ function plays an important role in the Ernst magnetization (\ref{magnetization}). 

Now, let us obtain the magnetized spacetime by using the Taub-NUT metric (\ref{metricTaubNUT}) as the seed solution. Following (\ref{magnetization}), the corresponding magnetized Ernst gravitational potential from (\ref{seedE}) and (\ref{Lambda}) can be written as
\be \label{Emag}
{\cal E}' = \frac{6Mrl{x}^{2}-{r}^{2}l+{l}^{3}-3l{r}^{2}{x}^{2}+3{l}^{3}{x}^{2} + i\left(\Delta_x \left\{r^3 + 3rl^2\right\} +2M{l}^{2}{x}^{2}\right)}{\delta_x b^2 {l}^{3}+ \left(1-{b}^{2}{r}^{2}\delta_x+6{b}^{2}Mr{x}^{2} \right) l + i\left\{\left( 3{b}^{2}r\Delta_x+2{b}^{2}M{x}^{2} \right) {l}^{2}+r+{b}^{2}{r}^{3}\Delta_x\right\}}\,.
\ee 
On the other hand, the resulting electromagnetic Ernst potential simply reads
\be \label{Phimag}
\Phi ' = -b {\cal E}'\,.
\ee  
This is obvious from (\ref{magnetization}) since the seed metric (\ref{metricTaubNUT}) has no associated electromagnetic Ernst potential, i.e. $\Phi = 0$. Consequently, the magnetized metric function
\be 
f' = {\mathop{\rm Re}\nolimits} \left\{ {{\cal E}'} \right\} - \left| {\Phi '} \right|^2 
\ee 
which is related to the seed function $f$ as $f' = \left| \Lambda \right|^{-2} f $ can be written as 
\be \label{fprimed}
f' = \frac{{\Delta _x \left( {r^2  + l^2 } \right)^2  - 4\Delta _r l^2 x^2 }}{\Xi} \,.
\ee 
In the last equation, we have used $\Xi = d_6 l^6 + d_4 l^4 +d_2 l^2 + d_0$ where
\[d_0 ={r}^{2} \left( 1+{r}^{2}{b}^{2}\Delta_x \right) ^{2}~~,~~d_6 = b^4 \delta_x^2\,,\]
\[d_4 = b^{2} \left( 7{r}^{2}{b}^{2}+24{b}^{2}{x}^{4}Mr+24{b}^{2}Mr{x}^{2}+6{x}^{2}+2-30{r}^{2}{b}^{2}{x}^{2}+4{b}^{2}{M}^{2}{x}^{4}-9{b}^{2}{r}^{2}{x}^{4} \right)\,, \]
and
\[d_2 = 1+ \left( 36{M}^{2}{r}^{2}{x}^{4}-40{r}^{3}{x}^{4}M-6{r}^{4}{x}^{2}+15{x}^{4}{r}^{4}+7{r}^{4}-8{r}^{3}M{x}^{2} \right) {b}^{4} + \left( 16Mr{x}^{2} 4{r}^{2}-12{r}^{2}{x}^{2} \right) {b}^{2}\,.
\]

Note that the new twist potential $\Psi'$ is associated to the transformed Ernst potential ${\cal E}'$ as
\be \label{Psiprimed}
\Psi ' = \frac{-2l \left( {r}^{3}+r{l}^{2}+{r}^{3}{x}^{2}-3r{l}^{2}{x}^{2}+M{l}^{2}{x}^{2}-3M{r}^{2}{x}^{2} \right) }{\Xi}\,.
\ee 
Furthermore, integrating out (\ref{drwp}) and (\ref{dxwp}) gives us
\be 
{\omega '}= \frac{2lx\Delta_r \left\{c_4 l^4 + c_2 l^2 +c_0\right\} }{{\Delta _x \left( {r^2  + l^2 } \right)^2  - 4\Delta _r l^2 x^2 }} \,,
\ee 
where
\be 
c_4 = -b^4 \delta_x \Delta_x\,,
\ee 
\be 
c_2 = 2{b}^{4} \left(3{r}^{2}{x}^{4} -2{x}^{4}Mr+2{M}^{2}{x}^{4}+3
{r}^{2}-6{r}^{2}{x}^{2}+2{x}^{2}Mr \right) \,,
\ee 
and
\be 
c_0 = 1+b^4 r^3 \Delta_x \left(rx^2+3r-4Mx^2\right)\,.
\ee 
Obviously, $\omega '$ reduces to (\ref{w0}) as one considers the limit $b\to0$. Using the obtained $\omega '$ and $f'$ functions above, the metric after magnetization now becomes
\be \label{metric.magnetized}
ds^2  =\frac{1}{f'} \left\{-{\Delta _r \Delta _x } dt^2 + \left( {r^2  + l^2 } \right)\left( {\frac{{dr^2 }}{{\Delta _r }} + \frac{{dx^2 }}{{\Delta _x }}} \right)\right\} +  f'\left( {d\phi  - \omega 'dt} \right)^2 \,.
\ee 

On the other hand, the accompanying vector field in solving the Einstein-Maxwell equations (\ref{eq.Einstein-Maxwell})
can be obtained from the electromagnetic Ernst potential $\Phi ' = A_{\phi} + i {\tilde A}_\phi$, where the vector component $A_t$ can be found after integrating (\ref{drAt}) and (\ref{dxAt}). Explicitly, these vector components read
\[
A_\phi = -\frac{b}{\Xi} \left\{ b^2 \Delta_x r^6 + \left(1+15{b}^{2}{l}^{2}{x}^{4}-{x}^{2}-6{b}^{2}{l}^{2}{x}^{2}+7{b}^{2}{l}^{2}
\right) r^4 -8{b}^{2}M{l}^{2}{x}^{2} \left( 5{x}^{2}+1 \right) r^3 \right.
\] 
\[
+{l}^{2}\left(2-6{x}^{2}-30{b}^{2}{l}^{2}{x}^{2}+36{b}^{2}{M}^{2}{x}^{4}+7{b}^{2}{l}^{2}-9{b}^{2}{l}^{2}{x}^{4} \right) r^2 + 8M{l}^{2}{x}^{2} \left(1+3{b}^{2}{l}^{2}\left\{1+x^2\right\}\right) r
\]
\be \label{Apmag}
\left. +{l}^{4}\left(4{b}^{2}{M}^{2}{x}^{4}+6{b}^{2}{l}^{2}{x}^{2}+3{x}^{2}+{b}^{2}{l}^{2}+1+9{b}^{2}{l}^{2}{x}^{4} \right) \right\}
\ee 
and
\[
A_t = -\frac{2lbx\Delta_r}{\Xi} \left\{ b^4 \Delta_x \left(3+x^2\right) r^4 -4b^4 x^2 M \Delta_x r^3 + 2{b}^{2} \left( 1+ x^2 + 3 b^2 \Delta_x^2 \right) r^2 
 \right.
\] 
\be \label{Atmag}
\left. -4b^2 x^2 M \left(1-b^2 l^2 \Delta_x\right) r + 4 l^2 b^4 M^2 x^4 -  \left(1 + {b}^{2}{l}^{2} \Delta_x \right)  \left( 3{b}^{2}{l}^{2}{x}^{2}+1+{b}^{2}{l}^{2} \right)   \right\}\,.
\ee 
It can be verified that this vector solution obeys the source-free condition, $\nabla_\mu F^{\mu \nu} = 0$. Moreover, considering the massless limit of (\ref{metric.magnetized}), (\ref{Apmag}), and (\ref{Atmag}) gives us the Melvin-Taub-NUT universe\footnote{The solution is given in appendix \ref{app.MelvinTaubNUT}.}, i.e. the Taub-NUT extension of the Melvin magnetic universe discovered in \cite{Melvin:1963qx,Griffiths:2009dfa}.

\begin{figure}
	\centering
	\includegraphics[scale=0.4]{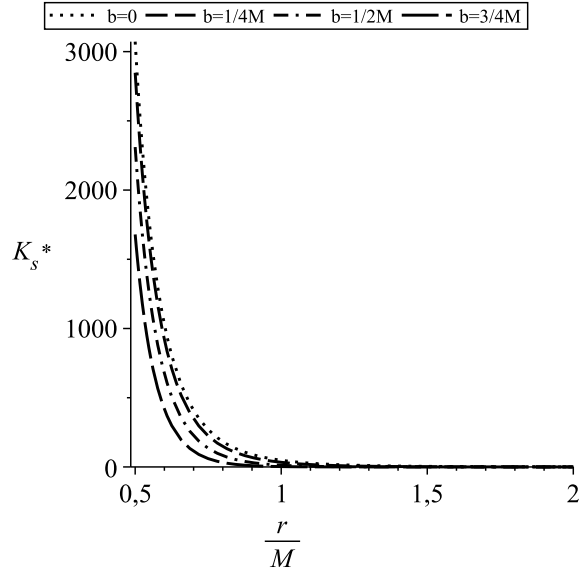}
	\caption{Some numerical evaluations for the Kretschmann scalar at equator in the absence of NUT parameter, where $K_s ^* = M^{-4} R_{\mu\nu\alpha\beta}R^{\mu\nu\alpha\beta}$.}\label{fig.Kr2l0}
\end{figure}

\begin{figure}
	\centering
	\includegraphics[scale=0.4]{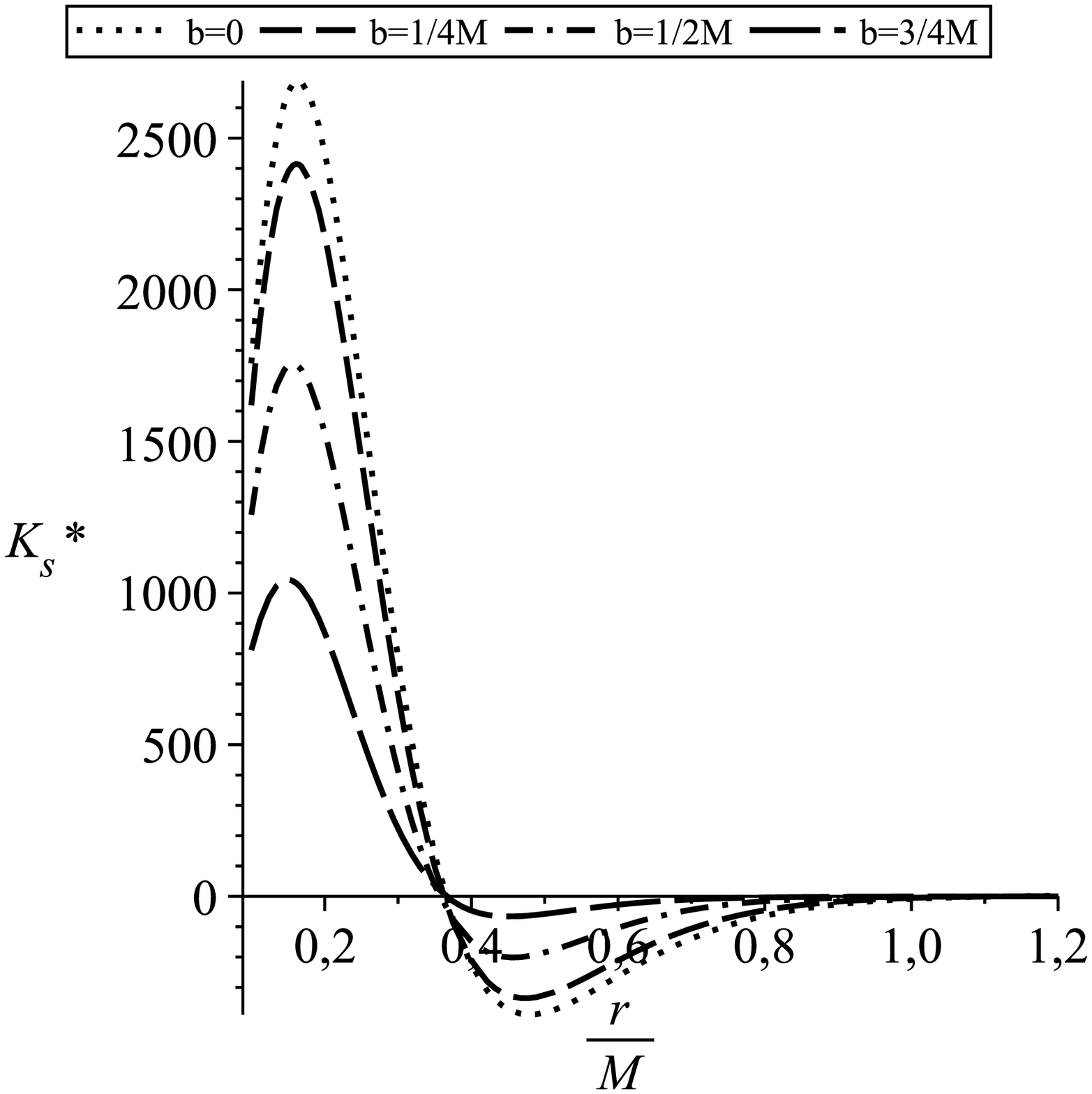}
	\caption{Some numerical evaluations for the Kretschmann scalar in the magnetized Taub-NUT spacetime with $l=M/2$ at equator, where $K_s ^* = M^{-4} R_{\mu\nu\alpha\beta}R^{\mu\nu\alpha\beta}$. }\label{fig.Kr2l05}
\end{figure}

\begin{figure}
	\centering
	\includegraphics[scale=0.4]{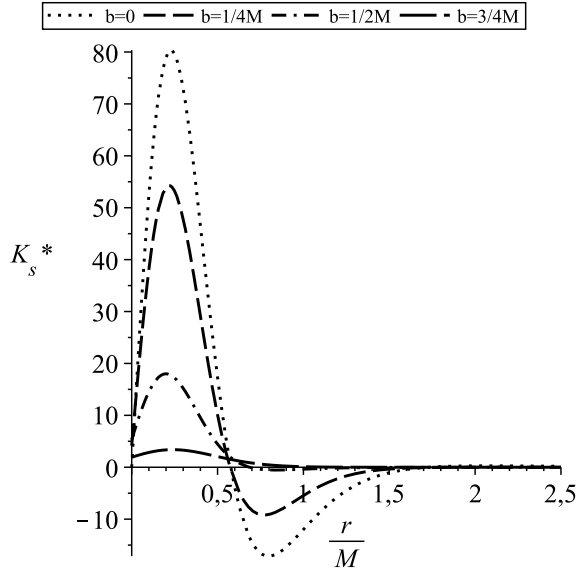}
	\caption{Some numerical evaluations for the Kretschmann scalar in the magnetized Taub-NUT spacetime with $l=M$ at equator, where $K_s ^* = M^{-4} R_{\mu\nu\alpha\beta}R^{\mu\nu\alpha\beta}$. }\label{fig.Kr2}
\end{figure}

\section{Surface geometry and closed timelike curve}\label{sec.4.SurfaceGeom}

In this section, let us study the horizon surface deformation and the closed timelike curve in the spacetime. Before we discuss these features, let us examine the Kretschmann scalar in spacetime, to justify the existence of the true singularity in the spacetime. However, the complexity of the spacetime solution (\ref{metric.magnetized}) hinders us to express the Kretschmann scalar explicitly. Therefore, we will perform some numerical evaluations and see whether the Kretschmann scalar can be singular at the origin. As we have mentioned in the introduction, spacetime with a NUT parameter has the conical singularity instead of a true one at the origin. This is known from the fact that typically spacetime with NUT parameter has a non-singular Kretschmann scalar at $r = 0$. In the absence of NUT parameter, the Kretschmann scalar blows up at origin even in the presence of an external magnetic field as depicted in fig. \ref{fig.Kr2l0}. However, the typical plots of Kretschmann scalar for a spacetime with NUT parameter appear in figs. \ref{fig.Kr2l05} and \ref{fig.Kr2}, which allow us to infer that the magnetized Taub-NUT spacetime does not possess a true singularity at the origin.  

As one would expect for a magnetized spacetime obtained by the Ersnt method, the existence of an external magnetic field does not affect the radius of the event horizon. It is the zero of $\Delta_r$ which happens to be the same as in the non-magnetized one, namely $r_+ = M + \sqrt{M^2 +l^2}$. Furthermore, the total area of horizon reads
\be \label{Area}
A = \int\limits_0^{2\pi } {\int\limits_{ - 1}^1 {\sqrt {{g_{xx}}{g_{\phi \phi }}} dxd\phi } }  = 4\pi \left( {r_ + ^2 + {l^2}} \right)\,,
\ee 
which is equal to the area of the generic Taub-NUT black hole. Consequently, the entropy of a magnetized Taub-NUT black hole will be the same as that of a non-magnetized one, namely $S = A/4$. 

However, the external magnetic field can distort the horizon of black hole, as reported in \cite{Wild:1980zz}. In getting to this conclusion, one can study the Gaussian curvature $ K = \tfrac{1}{2} R$ of the two dimensional surface of the horizon, where $R$ is the scalar curvature. For the magnetized Taub-NUT black hole, the corresponding two dimensional surface of horizon reads
\be \label{metric.2d.hor}
ds_{{\rm{hor}}}^2 = {{\left( {r_ + ^2 + {l^2}} \right)d{x^2}} \over {{f_ + }'{\Delta _x}}} + {f_ + }'d{\phi ^2}\,,
\ee 
where $f_+ '$ is $f'$ evaluated at $r_+$. Furthermore, the Gaussian curvature at equator can be found as
\[
{K_{x = 0}} = {{\left\{ {3{b^4}{l^8}} + 4r_+b^4\left(4M-9r_+\right)l^6 - \left( 2{r}_+^{4}{b}^{4}+32{r}_+^{3}M{b}^{4}+3 \right) {l}^{4} \right.} \over {{{\left( {{r_+^2} + {l^2}} \right)}^3}\left( {{b^4}{l^4} + 2{l^2}{b^2}\left\{ {1 + 3{b^2}{r_+^2}} \right\} + {{\left\{{1 + {b^2}{r_+^2}}\right\}}^2}} \right)}}
\]
\be \label{GaussianCURV}
\left. -2r_+ \left(8{b}^{4}M{r}_+^{4} -2{r}_+^{5}{b}^{4}-3r_++4M \right) {l}^{2}
+{{r_+^4}\left(1-b^4 r_+^4\right)} \right\}\,.
\ee 
Taking the limit $l\to 0$ from eq. above, we recover the Gaussian curvature of horizon at $x=0$ for \sch black hole immersed in a magnetic field \cite{Wild:1980zz}
\be \label{Kl0x0mag}
{K_{l=0,x = 0}} = {{1 - 4{b^2}{M^2}} \over {4{M^2}{{\left( {1 + {b^2}{M^2}} \right)}^3}}}\,.
\ee 
Note that in the absence of external magnetic field, this curvature takes the form
\be
{K_{{\rm{Taub - NUT}},x = 0}} = {1 \over {2{r_ + }\sqrt {{M^2} + {l^2}} }}\,,
\ee 
which is always positive just like in the \sch case. In the presence of magnetic field, the scalar curvature at horizon vanishes for $2bM = 1$, becomes negative for $2bM >1$, and positive for $2bM < 1$.  Particularly, the zero equatorial scalar curvature occurs for the magnetic field strength
\be \label{b0}
b = b_0 = \left\{ {\frac {r_+^2}{4 \left( 2M \left\{4{M}^{4}+6{l}^{2}{M}^{2}+3{l}^{4}\right\} \sqrt{M^2+l^2} +8{M}^{6}+16{l}^{2}{M}^{4}+11{l}^{4}{M}^{2}+2{l}^{6}\right)}}
\right\}^{\frac{1}{4}}\,.
\ee 

\begin{figure}
	\centering
	\includegraphics[scale=0.5]{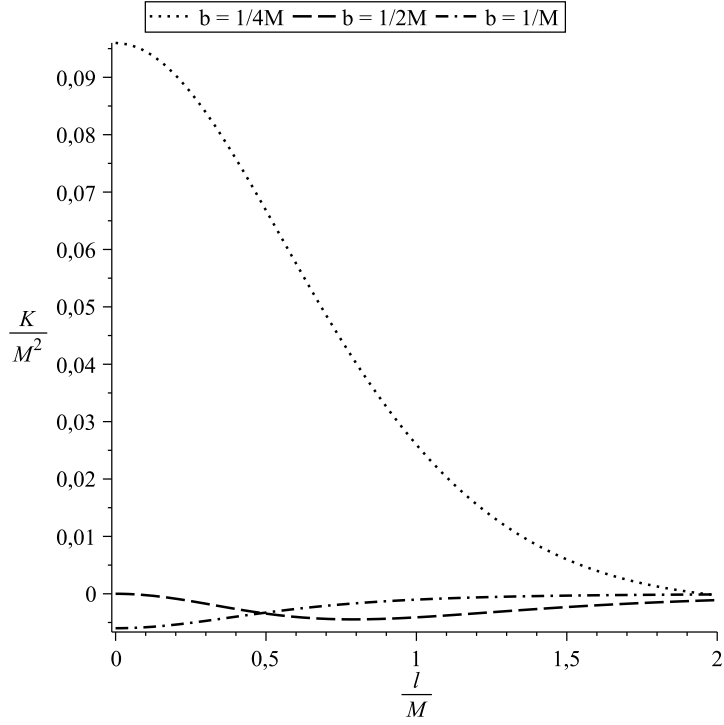}
	\caption{Gaussian curvature (\ref{GaussianCURV}) evaluated for some magnetic field strength $b$.}\label{fig.Kvsl}
\end{figure}

\begin{figure}
	\centering
	\includegraphics[scale=0.5]{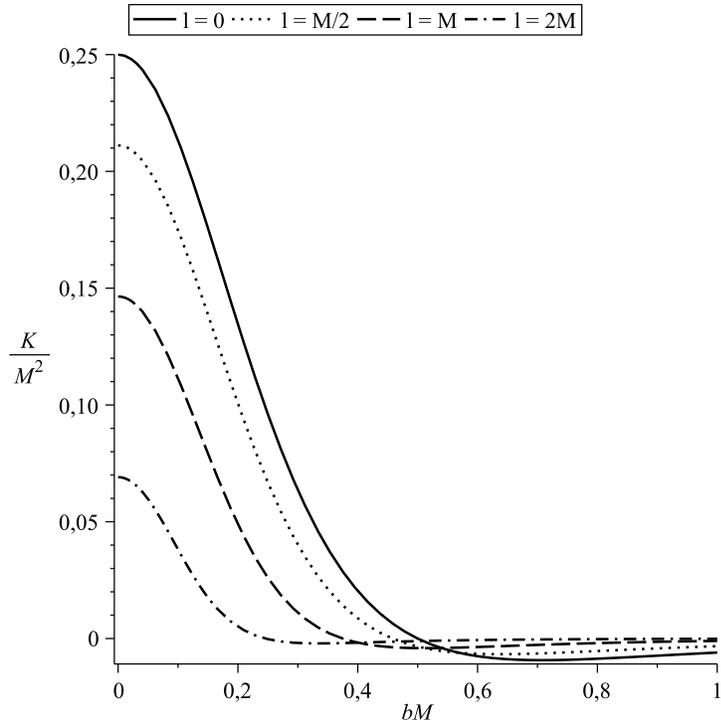}
	\caption{Gaussian curvature (\ref{GaussianCURV}) evaluated for some values of NUT parameter $l$.}\label{fig.Kvsb}
\end{figure}

Obviously, if we set $l\to 0$ in (\ref{b0}), we recover the condition for zero scalar curvature at horizon for the magnetized \sch black hole \cite{Wild:1980zz}. Furthermore, the positive curvature can be obtained for $b < b_0$, and the negative one for $b > b_0$. Nevertheless, the expression of (\ref{b0}) is quite complicated, so then we would evaluate eq. (\ref{GaussianCURV}) numerically to see how the curvature varies with $b$ and $l$. The case of null NUT parameter is presented in figs. \ref{fig.Kvsl} and \ref{fig.Kvsb}, where we can learn that the curvature can be positive, negative, or zero values depending on the external magnetic field strength. This is in agreement to the results reported in \cite{Wild:1980zz}. Furthermore, the curvature (\ref{GaussianCURV}) vanishes as $l\to \infty$ which is in agreement to the plots presented in fig. \ref{fig.Kvsl}, and plots in fig. \ref{fig.Kvsb} shows how the curvature (\ref{GaussianCURV}) changes as the magnetic field increases. 

\begin{figure}
	\centering
	\includegraphics[scale=0.4]{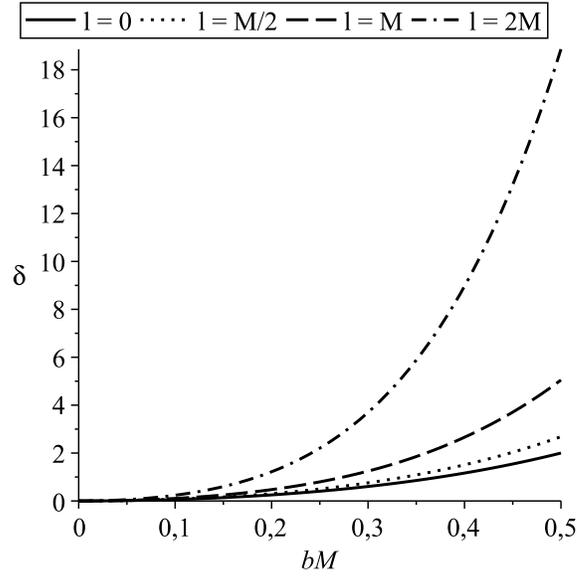}
	\caption{Plots of deviation from the spherical form of the surface as dictated by (\ref{delta}). We can observe that the deviation increases for the larger NUT parameter.}\label{fig.del}
\end{figure}

\begin{figure}
	\centering
	\includegraphics[scale=0.4]{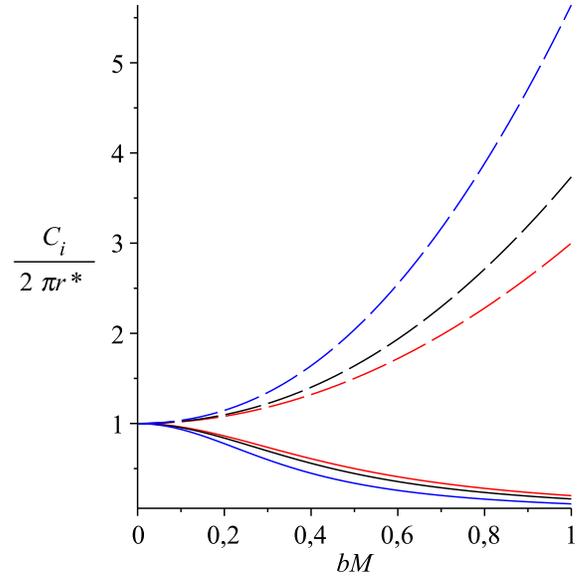}
	\caption{Solid lines are for subscript $i=e$ and dashed ones are for $i=p$. Here $r^* = r/M$ and the red, black, and blue colors denote the cases of $l=0$, $l=M/2$, and $l=M$, respectively. Here we learn that the gap between $C_p$ and $C_e$ gets bigger as the value of NUT parameter $l$ increases, confirming the results presented in fig. \ref{fig.del}.}\label{fig.C2pi}
\end{figure}

Another way to see how the external magnetic field deforms the horizon can be done by studying the shape of horizon as performed in \cite{Wild:1980zz} and \cite{Booth:2015nwa}. Surely, the generic \sch black hole horizon is a sphere. However, the horizon of a magnetized \sch black hole can form an oval shape, or even an hourglass appearance for a sufficiently large magnetic field \cite{Booth:2015nwa}. We show here that the horizon in the spacetime with NUT parameter also exhibits this effect. It can be understood from the previous finding where the equatorial Gaussian curvature can be negative for  $b > b_0$. To illustrate this prolateness effect, let us compute the equatorial circumference of horizon $C_e$, and the polar one $C_p$ as well. Since the integration would be in a full cycle, let us return to the standard Boyer-Lindquist type coordinate $\left(t,r,\theta,\phi\right)$. The standard textbook definition for these circumferences are
\be
{C_p} = \int\limits_0^{2\pi } {\sqrt {{g_{\theta \theta }}} d\theta } \,,
\ee 
and
\be 
{C_e} = {\left. {\int\limits_0^{2\pi } {\sqrt {{g_{\phi \phi }}} d\phi } } \right|_{\theta  = \pi /2}}\,.
\ee 
Following \cite{Wild:1980zz}, we can define the quantity $\delta$ which denotes the prolateness of horizon as a function of magnetic field $b$, 
\be \label{delta}
\delta  = {{{C_p} - {C_e}} \over {{C_e}}}\,.
\ee 
Note that, for the seed solution (\ref{metricTaubNUT}), this quantity vanishes implying the spherical horizon. In terms of $\gamma$ and $f'$ functions, expressed in $\theta$ instead of $x$, the corresponding metric functions incorporated in $C_p$ and $C_e$ are
\be 
g_{\theta\theta} = \frac{e^{2\gamma\left(r_+,\theta\right)}}{f'\left(r_+,\theta\right)}
~~{\rm and}~~
g_{\phi\phi} = f'\left(r_+,\tfrac{\pi}{2}\right)\,.
\ee 
To illustrate the shape of horizon, we provide numerical plots for some cases of NUT parameters in figs. \ref{fig.del} and \ref{fig.C2pi}.

\begin{figure}
	\centering
	\includegraphics[scale=0.4]{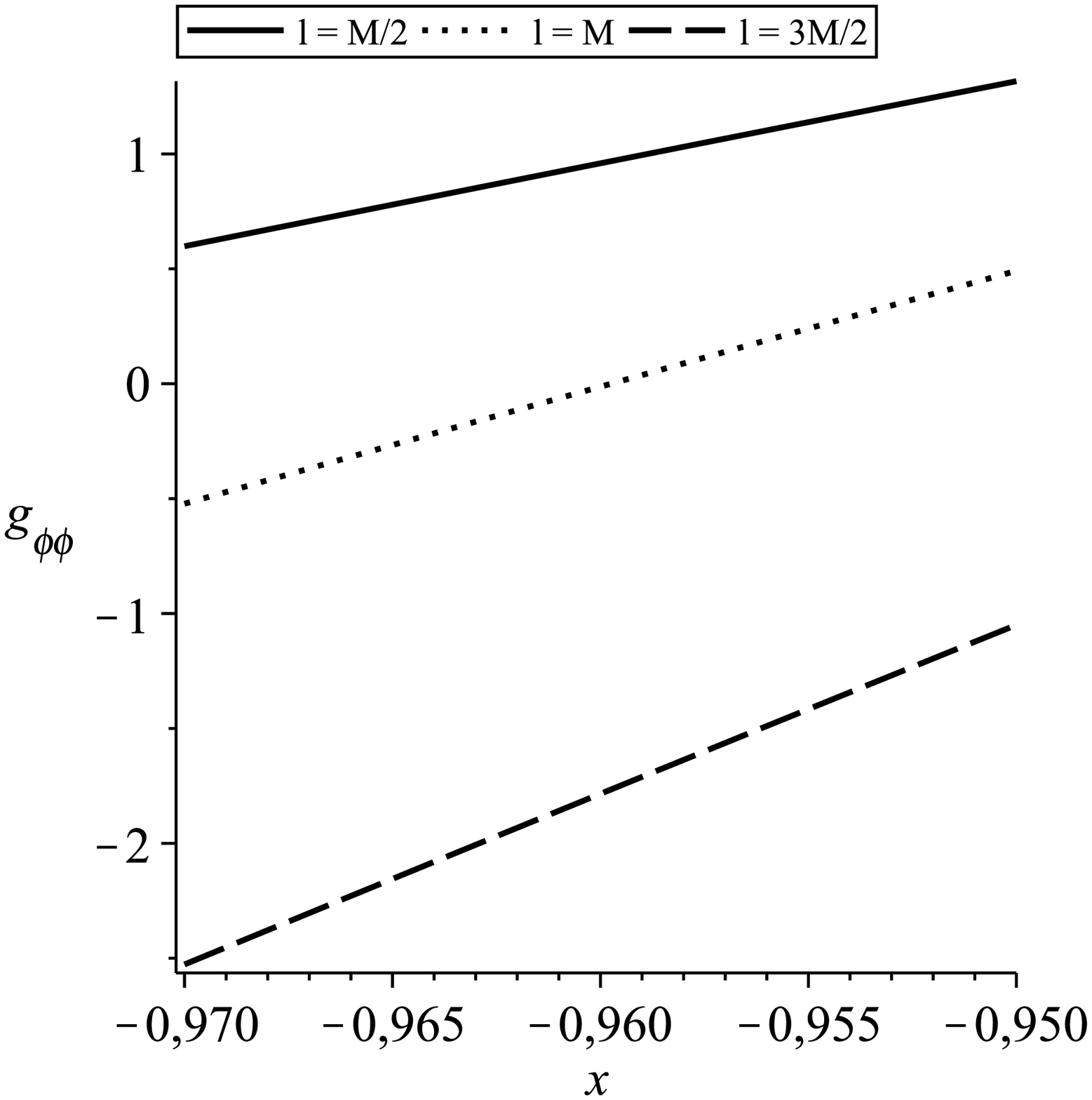}
	\caption{Evaluation of $g_{\phi\phi}$ in the absence of external magnetic field and for some values of $l$'s.}\label{fig.ggppl}
\end{figure}

\begin{figure}
	\centering
	\includegraphics[scale=0.4]{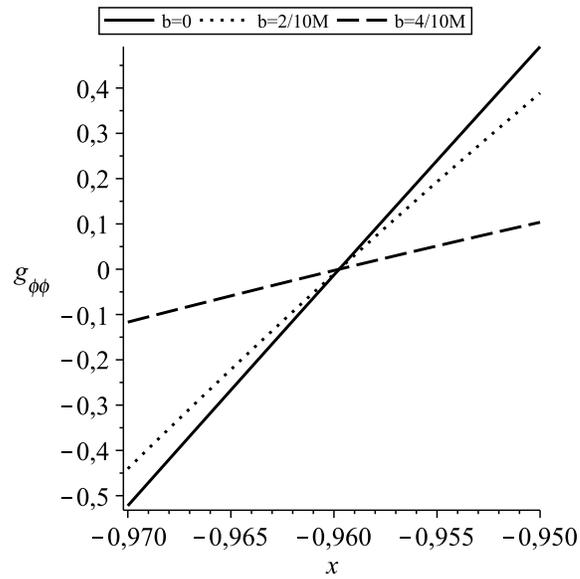}
	\caption{Evaluation of $g_{\phi\phi}$ for $l=M$ and some values of $b$, over the same angles as in fig. \ref{fig.ggppl}.}\label{fig.gppB}
\end{figure}

Now let us turn to the discussion of another aspect in the magnetized Taub-NUT spacetime, namely the existence of closed timelike curve (CTC). It is well known that the Taub-NUT spacetime possesses the CTC \cite{Griffiths:2009dfa} since $g_{\phi\phi}$ in the seed metric (\ref{metricTaubNUT}) becomes negative for
\be \label{timelike.cons.ph2}
x < \frac{4l^2 \Delta_r - \left(r^2+l^2\right)^2}{4l^2 \Delta_r + \left(r^2+l^2\right)^2}\,.
\ee 
Then it is natural to ask whether the CTC can also occur in the magnetized version of (\ref{metricTaubNUT}). If it exists, then how would the external magnetic field influence the existing CTC? Related to the magnetized line element, the $\left(\phi,\phi\right)$ component of the metric changes in the form
\be 
{g_{\phi \phi }} \to g{'_{\phi \phi }} = {\left| \Lambda  \right|^{ - 2}}{g_{\phi \phi }}\,.
\ee 
Clearly, it is troublesome to express the condition for CTC occurrence in the magnetized Taub-NUT spacetime analogous to eq. (\ref{timelike.cons.ph2}) of the non-magnetized one. Therefore, we provide figs. \ref{fig.ggppl} and \ref{fig.gppB} which show some numerical evaluations of $g_{\phi\phi}$ for some particular $b$ and $l$. Curves in fig. \ref{fig.ggppl} are some results in the absence of magnetic field where the CTC can be understood to exist from the negative value of $g_{\phi\phi}$. On the other hand, plots in fig. \ref{fig.gppB} associate to the case with magnetic field which show that CTC can occur as well.

\section{Electromagnetic properties}\label{sec.EM}

In this section, let us study some properties of electromagnetic fields around the magnetized Taub-NUT black hole. The electric and magnetic fields are given by
\be \label{Efield}
E_\alpha   =  - F_{\alpha \beta } u^\beta  \,,
\ee 
and
\be \label{Bfield}
B_\alpha   = \frac{1}{2}\varepsilon _{\alpha \beta \mu \nu } F^{\mu \nu } u^\beta  \,,
\ee 
respectively, where $u^\alpha = [1,0,0,0]$ is the stationary Killing vector. For the solutions presented in eqs. (\ref{metric.magnetized}), (\ref{Apmag}), and (\ref{Atmag}), the non-zero components of electric and magnetic fields are
\be\label{Er} 
E_r  = \frac{{4xbl\left( {f_{10} l^{10}  + f_8 l^8  + f_6 l^6  + f_4 l^4  + f_2 l^2  + f_0 } \right)}}{{\Xi _r^2 }}\,,
\ee
and
\be \label{Ex}
E_x  = \frac{{2 b l \left(r^2 - 2Mr -l^2\right)\left( {h_{10} l^{10}+h_{8} l^{8}  + h_6 l^6  + h_4 l^4  + h_2 l^2  + h_0 } \right)}}{{\Xi _x^2 }}\,.
\ee 
On the other hand, the non-zero components of magnetic field are
\be \label{Br}
B_r  = -\frac{{2xb\left( {{\tilde f}_{12} l^{12} + {\tilde f}_{10} l^{10}  + {\tilde f}_8 l^8  + {\tilde f}_6 l^6  + {\tilde f}_4 l^4  + {\tilde f}_2 l^2  + {\tilde f}_0 } \right)}}{{{\Upsilon} _r^3 }}\,,
\ee 
and
\be \label{Bx}
B_x  = -\frac{{2 b \left(r^2- 2Mr-l^2\right) \left( {{\tilde h}_{10} l^{10}+{\tilde h}_{8} l^{8}  + {\tilde h}_6 l^6  + {\tilde h}_4 l^4  + {\tilde h}_2 l^2  + {\tilde h}_0 } \right)}}{{\Upsilon_x^3 }}\,.
\ee 
The $r$ and $x$ dependent functions $f_j$, $h_j$, ${\tilde f}_j$, ${\tilde h}_j$, $\Xi_r$, $\Xi_x$, ${\Upsilon}_r$, and ${\Upsilon}_x$ above are provided in the appendix \ref{app.Functions}. Furthermore, the asymptotic of each component above can be found as
\be 
\mathop {\lim }\limits_{r \to \infty } E_r  = 0~~,~~\mathop {\lim }\limits_{r \to \infty } B_r  = 0~~,~~\mathop {\lim }\limits_{r \to \infty } B_x  = 0\,,
\ee 
and
\be 
\mathop {\lim }\limits_{r \to \infty } E_x  = {\frac {2lb \left( {x}^{4}-6{x}^{2}-3 \right) }{1-2{x}^{2}+{x}^{4}}}\,.
\ee 
 
\begin{figure}
	\centering
	\includegraphics[scale=0.4]{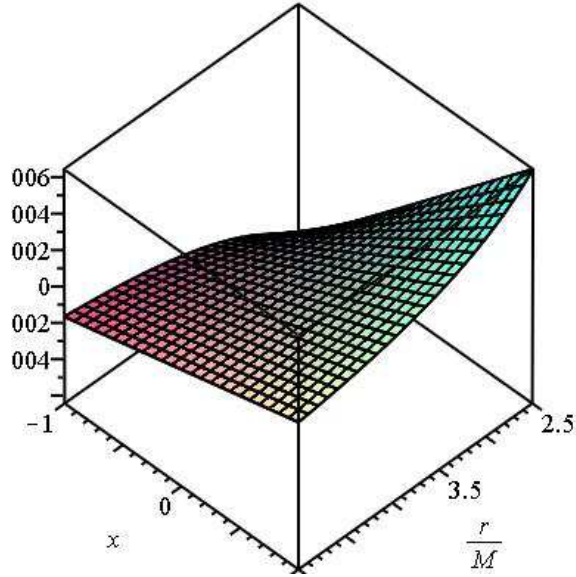}
	\caption{Plot of dimensionless $E_r^* = ME_r$ for $b = 0.01 M$ and $l=0.1 M$ in the range of $2.5 M \le r \le 5M$ and $-1 \le x \le 1$.}\label{fig.Erb001l01}
\end{figure}

\begin{figure}
	\centering
	\includegraphics[scale=0.4]{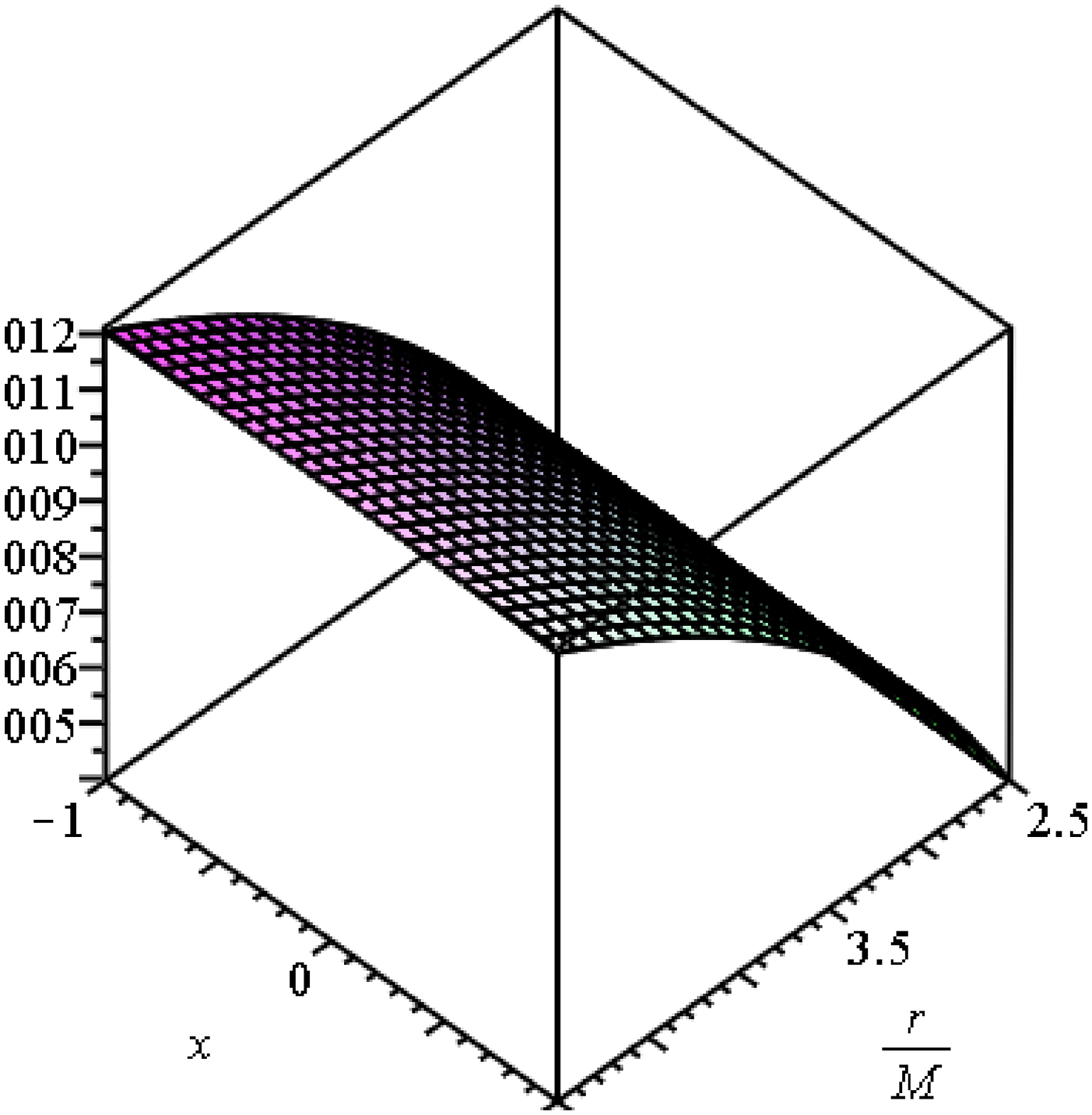}
	\caption{Plot of dimensionless $E_x^* = E_x$ for $b = 0.01 M$ and $l=0.1 M$ in the range of $2.5 M \le r \le 5M$ and $-1 \le x \le 1$.}\label{fig.Exb001l01}
\end{figure}

\begin{figure}
	\centering
	\includegraphics[scale=0.4]{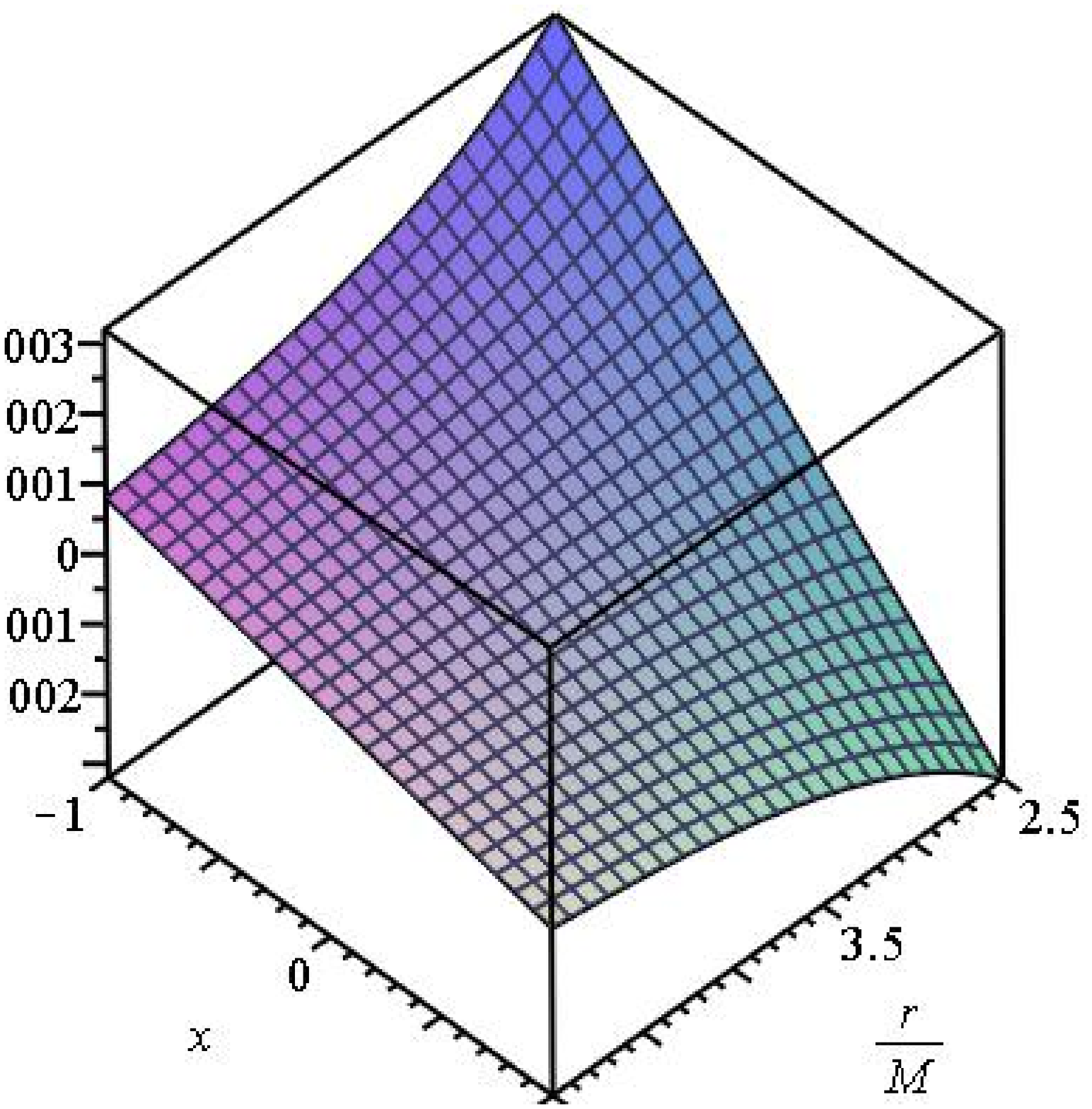}
	\caption{Plot of dimensionless $B_r^* = M^3 E_r$ for $b = 0.01 M$ and $l=0.1 M$ in the range of $2.5 M \le r \le 5M$ and $-1 \le x \le 1$.}\label{fig.Brb001l01}
\end{figure}

\begin{figure}
	\centering
	\includegraphics[scale=0.4]{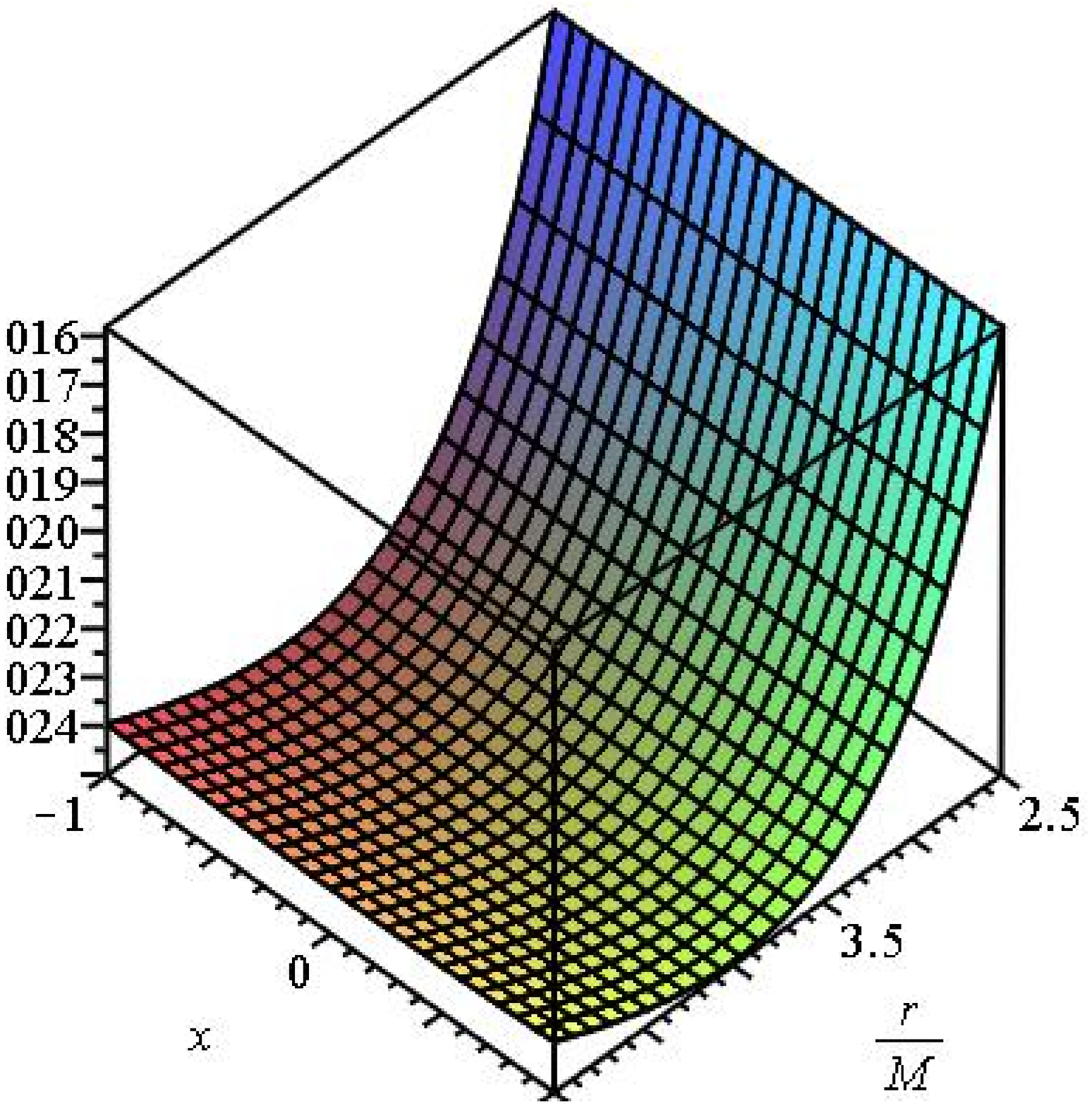}
	\caption{Plot of dimensionless $B_x^* = M B_x$ for $b = 0.01 M$ and $l=0.1 M$ in the range of $2.5 M \le r \le 5M$ and $-1 \le x \le 1$.}\label{fig.Bxb001l01}
\end{figure}

Obviously, the lengthy expressions for the electric and magnetic fields (\ref{Er}), (\ref{Ex}), (\ref{Br}), and (\ref{Bx}) hinder us to extract some more qualitative results related to the fields properties around the black hole. However, the behavior of these fields can still be studied by using some numerical examples as appear in figs. \ref{fig.Erb001l01}, \ref{fig.Exb001l01}, \ref{fig.Brb001l01}, \ref{fig.Bxb001l01}. From fig. \ref{fig.Erb001l01}, we can learn that the magnitude of $E_r$ increases as going from $x=-1$ to $x=1$. On the other hand, from fig. \ref{fig.Exb001l01}, $E_x$ increases in magnitude for a larger $r$. Moreover, for $B_r$ as depicted in fig. \ref{fig.Brb001l01}, we learn that the maximum value for the considered numerical setups is at $x=-1$ and near horizon. 

\begin{figure}
	\centering
	\includegraphics[scale=0.4]{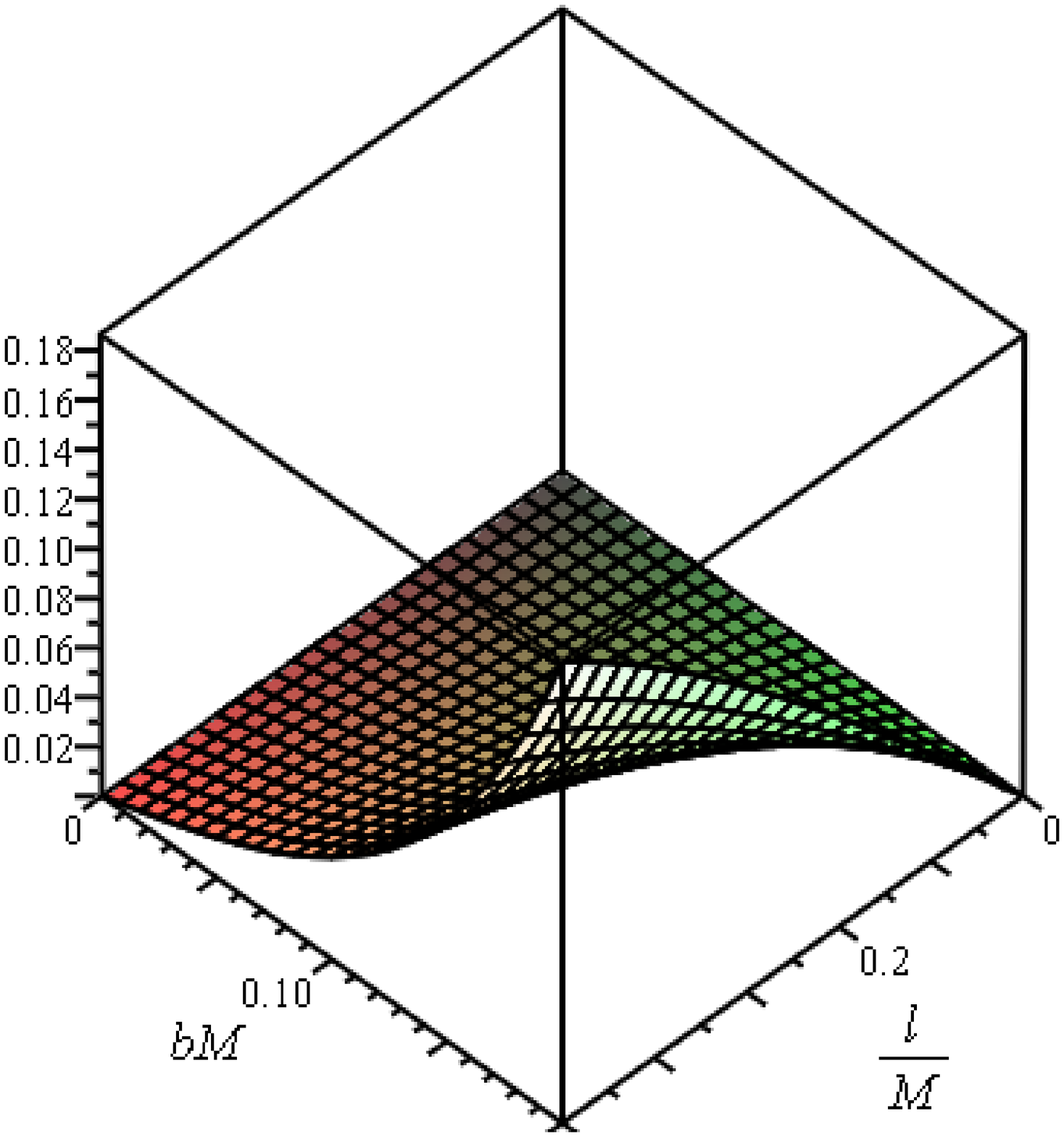}
	\caption{Equatorial $E_x$ at $r=3 M$ for $0 \le l \le 0.5M$ and $0 \le b \le 0.2/M$.}\label{fig.Exr3x0}
\end{figure}

\begin{figure}
	\centering
	\includegraphics[scale=0.4]{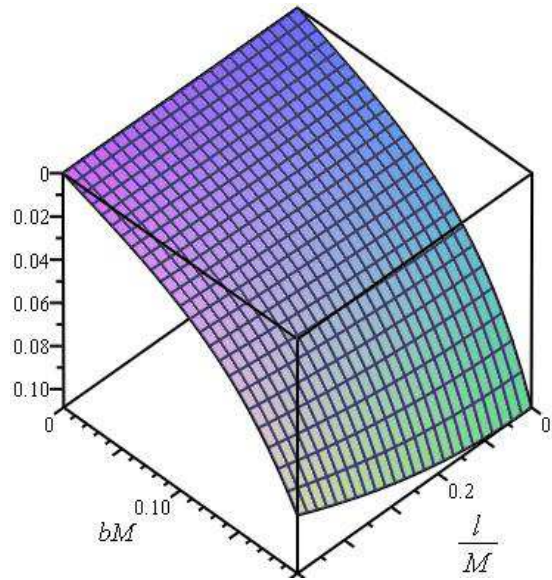}
	\caption{Equatorial dimensionless $B_x^*$ at $r=3 M$ for $0 \le l \le 0.5M$ and $0 \le b \le 0.2/M$.}\label{fig.Bxr3x0}
\end{figure}

Now let us evaluate these fields at equator. Obviously the $r$-component of the fields vanish there, while the $x$-components evaluated at $r=3M$ are depicted in figs. \ref{fig.Exr3x0} and \ref{fig.Bxr3x0}. From \ref{fig.Exr3x0} we observe that $E_x$ increases as $b$ and $l$ grow, and from fig. \ref{fig.Bxr3x0} we learn that the magnitude grows as magnetic field increases. However, the presence of NUT parameter leads to the smaller magnitude of $B_x$ as illustrated in fig. \ref{fig.Bxr3x0}. 

\section{Conclusion}

In this paper, we have presented a novel solution in Einstein-Maxwell theory namely the Taub-NUT extension of magnetized black hole reported in \cite{Ernst}. To get the solution, we have employed the Ernst magnetization to the Taub-NUT spacetime. Typical for a spacetime with NUT parameter, the equatorial Kretschmann scalar does not blow up at the origin. Moreover, we find that the black hole surface in magnetized Taub-NUT spacetime deforms due the presence of an external magnetic field, similar to the magnetized \sch case reported in \cite{Wild:1980zz}. Furthermore, the existence of NUT parameter leads to the occurrence of a closed timelike curve in the spacetime as shown in fig \ref{fig.gppB}.  In addition to the surface geometry and closed timelike curve discussions, we also add some studies on the electromagnetic properties in the spacetime. 

There are several interesting future problems that we can investigate for the new spacetime solution presented in this work, for example the conserved charges and thermodynamical related. In particular, it is associated with the Melvin-Taub-NUT solution \cite{Melvin:1963qx} as given in the appendix. The distribution of energy in Melvin spacetime had been studied in \cite{Xulu:1999gf,Xulu:1999tx}, and its stability against some perturbations was investigated in \cite{Thorne:1965}. Investigating the energy distribution associate to the Melvin-Taub-NUT spacetime and its stability against some perturbations are worth consideration.

\section*{Acknowledgement}

This work is supported by Lembaga Penelitian Dan Pengabdian Kepada Masyarakat, Parahyangan Catholic University. I thank Merry K. Nainggolan for her support and encouragement.

\appendix
\section{Melvin Taub NUT spacetime}\label{app.MelvinTaubNUT}

In this appendix we provide the solution describing the Taub-NUT extension of the Melvin magnetic universe. The metric components are  
\be 
g_{rr}  = \frac{\Upsilon }{{r^2  - l^2 }} = \frac{{\Delta _x g_{xx} }}{{r^2  - l^2 }}\,,
\ee 
\[
g_{tt} = {\Upsilon}^{-1} \left\{  \left(1+b^2 r^2 \Delta_x\right)^4 + b^8 \Delta_x \delta_x \left(5x^4 + 10x^2 + 1\right) l^8 - 4 b^6 \Delta_x \left(15{b}^{2}{r}^{2}{x}^{4}\left\{1+x^2\right\} -9{x}^{4} \right.  \right.
\]
\[ \left. +21{x}^{2}{b}^{2}{r}^{2} -6{x}^{2}-3{b}^{2}{r}^{2}-1\right) l^6 +2b^4 \Delta_x \left(5{b}^{4}{x}^{6}{r}^{4}+105{b}^{4}{r}^{4}{x}^{4}-18{r}^{2}{b}^{2}
{x}^{4} \right. \]
\[
\left. -33{b}^{4}{r}^{4}{x}^{2}-60{x}^{2}{b}^{2}{r}^{2}+5{x}^{2}
+14{b}^{2}{r}^{2}+3+19{b}^{4}{r}^{4}\right) l^4 - 4b^2 \Delta_x \left(7{r}^{6}{b}^{6}{x}^{6}-15{b}^{4}{r}^{4}{x}^{4} \right.
\]
\be 
\left. \left. -{r}^{6}{b}^{6}{x}^{4}+13{r}^{6}{b}^{6}{x}^{2}+9{x}^{2}{b}^{2}{r}^{2}+6{b}^{4}{r}^{4
}{x}^{2}-5{b}^{2}{r}^{2}-3{r}^{6}{b}^{6}-7{b}^{4}{r}^{4}-1\right) l^2 \right\}\,,
\ee 
\be 
g_{t\phi} = \frac{2lx}{\Upsilon}\left(l^2-r^2\right) \left(1-{b}^{4}{r}^{4}{x}^{2}\left\{2+x^2\right\}+3{b}^{4}{r}^{4}+6l^2b^4r^2\Delta_x^2+3{l}^{4}{b}^{4}{x}^{4}-{l}^{4}{b}^{4}\left\{1+2x^2\right\}\right)\,,
\ee 
\be 
g_{\phi\phi} = \Upsilon^{-1} \left({r}^{4}-{r}^{4}{x}^{2}+2
{r}^{2}{l}^{2}+{l}^{4}-6{r}^{2}{l}^{2}{x}^{2}+3{l}^{4}{x}^{2}\right)\,,
\ee 
where
\[
\Upsilon = b^4 \delta_x^2 l^6  + \left\{1+ {b}^{2}{r}^{2} \left( 7{b}^{2}{r}^{2}+15{r}^{2}{b}^{2}{x}^{4}-6{x}^{2}{b}^{2}{r}^{2}+4-12{x}^{2} \right)\right\} l^2
\]
\be 
-b^2 \left(9{r}^{2}{b}^{2}{x}^{4}+30{x}^{2}{b}^{2}{r}^{2}-7{b}^{2}{r}^{2}-2
-6{x}^{2}\right) l^4 + r^2 \left(1+b^2 r^2 \Delta_x\right)^2\,,
\ee
with $\delta_x$ as introduced in (\ref{Lambda}). The associated vector components are
\[
A_t = 2lbx\left(r^2-l^2\right)\Upsilon^{-1} \left\{ 1+{b}^{4}{r}^{4}{x}^{4}+2{b}^{4}{r}^{4}{x}^{2}-3{b}^{4}{r}^{4} -6{l}^{2}{r}^{2}{b}^{4}{x}^{4}+12{l}^{2}{b}^{4}{r}^{2}{x}^{2} -2{x}^{2}{b}^{2}{r}^{2}\right.
\]
\be 
\left. -6{l}^{2}{b}^{4}{r}^{2}-2{b}^{2}{r}^{2}+2{l}^{4}
{b}^{4}{x}^{2}+2{b}^{2}{l}^{2}+{l}^{4}{b}^{4}+2{b}^{2}{l}^{2}{x}^{2}-3{l}^{4}{b}^{4}{x}^{4} \right\}\,,
\ee 
and
\[
A_\phi = -b \Upsilon^{-1} \left\{ {r}^{6}{x}^{4}{b}^{2}-6{r}^{4}{b}^{2}{l}^{2}{x}^{2}-30{r}^{2}{b}^{
	2}{l}^{4}{x}^{2}+{l}^{6}{b}^{2}+7{r}^{4}{b}^{2}{l}^{2}+7{r}^{2}{b}
^{2}{l}^{4}-2{r}^{6}{b}^{2}{x}^{2}+6{l}^{6}{b}^{2}{x}^{2}\right.
\]
\be 
\left. +9{b}^{2}{l}^{6}{x}^{4}-9{r}^{2}{l}^{4}{x}^{4}{b}^{2}+15{b}^{2}{l}^{2}{r}
^{4}{x}^{4}-6{r}^{2}{l}^{2}{x}^{2}+2{r}^{2}{l}^{2}-{r}^{4}{x}^{2}+
3{l}^{4}{x}^{2}+{r}^{4}+{l}^{4}+{r}^{6}{b}^{2} \right\}\,.
\ee 
This solution reduces to that of the Melvin universe \cite{Melvin:1963qx} as the limit $l\to 0$ is considered.

\section{Functions in the components of electric and magnetic fields}\label{app.Functions}

The followings are functions appearing in (\ref{Er})
\be 
f_{10} = {b}^{8} \Delta_x  \left( 1+3{x}^{2}
\right)  \left( 9M{x}^{4}-18r{x}^{4}+8M{x}^{2}-12{x}^{2}r-M+
14r \right) \,,
\ee 
\[ 
f_8 = -{b}^{6} \left( 10M{x}^{2}+24M{x}^{4}-25M{r}^{2}{b}^{2}+36r-8
{r}^{3}{b}^{2}-72{b}^{2}{M}^{2}r{x}^{4}-102M{r}^{2}{b}^{2}{x}^{2
}+168{b}^{2}r{x}^{8}{M}^{2}\right.
\]
\[
+72r{x}^{6}+232{b}^{2}M{r}^{2}{x}^{4}
-16{b}^{2}{r}^{3}{x}^{4}-495{b}^{2}{r}^{2}{x}^{8}M+24{b}^{2}{x}^
{6}{M}^{3}-30M{x}^{6}-108r{x}^{4}+216{b}^{2}{r}^{3}{x}^{8}
\]
\be
\left. -4M-
152{b}^{2}{x}^{6}{M}^{2}r+390{b}^{2}{r}^{2}{x}^{6}M-8{b}^{2}{M}^
{2}r{x}^{2}+8{b}^{2}{x}^{8}{M}^{3}-192{b}^{2}{r}^{3}{x}^{6}
\right) \,,
\ee
\[
f_6 = 2{b}^{4} \left( 39{b
}^{4}{r}^{4}M-284{b}^{4}{r}^{5}{x}^{4}-8{r}^{3}{b}^{2}-30M{r}^{2}{b}^{2}-24{b}^{2}{r}^{3}{x}^{4}+24{b}^{2
}{r}^{3}{x}^{6}+12{b}^{2}{x}^{6}{M}^{3}\right.
\]
\[
+128{b}^{4}{M}^{2}{r}^{3}{x
}^{2}+88{b}^{4}{M}^{3}{r}^{2}{x}^{4}-360{b}^{4}{M}^{2}{x}^{4}{r}^{
	3}-150{b}^{4}M{x}^{2}{r}^{4}+496{b}^{4}M{x}^{4}{r}^{4}-48M{r}^{2
}{b}^{2}{x}^{2}-4{b}^{2}{M}^{2}r{x}^{2}
\]
\[
-38{b}^{4}{r}^{5}-8{x}^{2
}r+16r-3M-36{b}^{4}{x}^{6}{M}^{3}{r}^{2}-96{b}^{4}{x}^{6}{M}^{2}{r}^{3}+16{b}^{4}{M}^{4}r{x}^{6}-96{b}^{4}{M}^{4}r{x}^{8}+264{	b}^{4}{x}^{8}{M}^{2}{r}^{3}
\]
\[
-231{b}^{4}{x}^{8}M{r}^{4}+108{b}^{4}{x}^{8}{M}^{3}{r}^{2}-154{b}^{4}{r}^{4}{x}^{6}M-24{b}^{2}{M}^{2}r{x}
^{4}+186{b}^{2}M{r}^{2}{x}^{4}+4{b}^{2}{x}^{6}{M}^{2}r-108{b}^{2
}{r}^{2}{x}^{6}M
\]
\be 
\left. -2M{x}^{4}+18{b}^{4}{r}^{5}{x}^{8}+192{b}^{4}{r}^{5}{x}^{6}+8{b}^{4}{M}^{5}{x}^{8}+8{b}^{2}{r}^{3}{x}^{2}+112{b}
^{4}{r}^{5}{x}^{2}+M{x}^{2} \right) 
\,,\ee 
\[
f_4 = -2{b}^{2} \left( 56{b
}^{6}{r}^{7}{x}^{4}+32{b}^{6}{r}^{7}{x}^{6}+58{b}^{4}{r}^{4}M-128
{b}^{6}{r}^{7}{x}^{2}-23M{r}^{2}{b}^{2}-48{b}^{4}{r}^{5}{x}^{4}-4{r}^{3}{b}^{2}\right.
\]
\[
+8{b}^{2}{r}^{3}{x}^{4}-
294{x}^{6}{b}^{6}M{r}^{6}+220{x}^{4}{b}^{6}M{r}^{6}-88{b}^{6}{M}
^{2}{r}^{5}{x}^{2}+112{b}^{4}{M}^{2}{r}^{3}{x}^{2}+24{b}^{4}{M}^{3}{r}^{2}{x}^{4}-272{b}^{4}{M}^{2}{x}^{4}{r}^{3}
\]
\[
-82{b}^{4}M{x}^{2}{r}^{4}+134{b}^{4}M{x}^{4}{r}^{4}-3M{r}^{2}{b}^{2}{x}^{2}+4{b}^{2
}{M}^{2}r{x}^{2}+150{b}^{6}{x}^{2}M{r}^{6}+120{b}^{6}{M}^{2}{x}^{6}{r}^{5}+196{b}^{6}{M}^{3}{r}^{4}{x}^{6}
\]
\[
+32{b}^{6}{M}^{3}{r}^{4}{x}^{4}-280{b}^{6}{M}^{2}{r}^{5}{x}^{4}+12{b}^{6}{r}^{7}{x}^{8}-40
{b}^{4}{r}^{5}+72{b}^{6}{M}^{5}{r}^{2}{x}^{8}-192{b}^{6}{M}^{4}{r}
^{3}{x}^{8}+216{b}^{6}{x}^{8}{M}^{2}{r}^{5}
\]
\[
-63{b}^{6}{x}^{8}M{r}^{6}-84{b}^{6}{x}^{8}{M}^{3}{r}^{4}+28{b}^{6}{r}^{7}+6r-2M-13{b}^{6}M{r}^{6}+24{b}^{4}{x}^{6}{M}^{3}{r}^{2}+144{b}^{4}{x}^{6}{M}
^{2}{r}^{3}-16{b}^{4}{M}^{4}r{x}^{6}
\]
\be
\left. -110{b}^{4}{r}^{4}{x}^{6}M+12
{b}^{2}{M}^{2}r{x}^{4}-14{b}^{2}M{r}^{2}{x}^{4}+36{b}^{4}{r}^{5}
{x}^{6}+12{b}^{2}{r}^{3}{x}^{2}+52{b}^{4}{r}^{5}{x}^{2}-M{x}^{2}
\right) 
\,,\ee 
\[
f_2 = 16{b}^{4}{r}^{5}{x}^{4}+32{b}^{6}{r}^{7}{x}^{4}+32{b}^{6}{r}^{7}
{x}^{6}-20{b}^{8}{r}^{9}{x}^{4}+22{b}^{8}{r}^{9}{x}^{8}+62{b}^{4}{r}^{4}M-96{b}^{6}{r}^{7}{x}^{2}-12M{r}^{2}{b}^{2}
\]
\[
+16{b}^{8}{r}
^{9}{x}^{2}+3{b}^{8}M{r}^{8}+128{b}^{8}{M}^{2}{r}^{7}{x}^{6}-168
{x}^{6}{b}^{6}M{r}^{6}-4{x}^{4}{b}^{6}M{r}^{6}+102{b}^{8}M{r}^{8}{x}^{2}+36{b}^{8}M{r}^{8}{x}^{4}
\]
\[
-200{b}^{6}{M}^{2}{r}^{5}{x}^{2}+32
{b}^{4}{M}^{2}{r}^{3}{x}^{2}-32{b}^{4}{M}^{3}{r}^{2}{x}^{4}+64{b
}^{4}{M}^{2}{x}^{4}{r}^{3}-74{b}^{4}M{x}^{2}{r}^{4}-60{b}^{4}M{x}^
{4}{r}^{4}-64{b}^{8}{M}^{2}{r}^{7}{x}^{2}
\]
\[
+336{b}^{8}{M}^{3}{r}^{6}
{x}^{4}-12M{r}^{2}{b}^{2}{x}^{2}+288{b}^{8}{M}^{4}{r}^{5}{x}^{8}-
288{b}^{8}{M}^{4}{r}^{5}{x}^{6}-6{b}^{8}{r}^{8}{x}^{6}M-135{b}^{8}{r}^{8}{x}^{8}M+336{b}^{8}{r}^{7}{x}^{8}{M}^{2}
\]
\[
+8{b}^{2}{M}^{2}r
{x}^{2}+192{b}^{6}{x}^{2}M{r}^{6}+264{b}^{6}{M}^{2}{x}^{6}{r}^{5}-
200{b}^{6}{M}^{3}{r}^{4}{x}^{6}-504{b}^{8}{M}^{3}{r}^{6}{x}^{8}+
176{b}^{6}{M}^{3}{r}^{4}{x}^{4}-400{b}^{8}{M}^{2}{r}^{7}{x}^{4}
\]
\be 
-48
{b}^{6}{M}^{2}{r}^{5}{x}^{4}+168{b}^{8}{M}^{3}{r}^{6}{x}^{6}-32{b}^{4}{r}^{5}+32{b}^{6}{r}^{7}-18{b}^{8}{r}^{9}+2r-M-20{b}^{6}
M{r}^{6}+8{b}^{2}{r}^{3}{x}^{2}+32{b}^{4}{r}^{5}{x}^{2}
\,,\ee
\be 
f_0 = 
-{r}^{2} \left( 1+{b}^{2}{r}^{2}{x}^{2}-{b}^{2}{r}^{2} \right) ^{2}
\left( 3{b}^{4}{r}^{4}M+8{b}^{4}{M}^{2}{x}^{4}{r}^{3}-8{b}^{4}{M}^{2}{r}^{3}{x}^{2}-4{r}^{3}{b}^{2}+6M
{r}^{2}{b}^{2}-M -3{b}^{4}M{x}^{4}{r}^{4}\right) 
\,,\ee 
\[
\Xi_r = \left\{ 12{b}^{2}{l}^{2}{r}^{2}{x}^{2}-
4{b}^{2}{l}^{2}{r}^{2}-6{b}^{2}{l}^{4}{x}^{2}+2{b}^{2}{r}^{4}{x}
^{2}-2{b}^{2}{l}^{4}-2{b}^{2}{r}^{4}-2{b}^{4}{r}^{6}{x}^{2}+{b}^
{4}{r}^{6}{x}^{4}+9{l}^{6}{b}^{4}{x}^{4}\right.
\]
\[
-16{b}^{2}{l}^{2}Mr{x}^{2}+6{l}^{6}{b}^{4}{x}^{2}+7
{l}^{4}{b}^{4}{r}^{2}+7{l}^{2}{b}^{4}{r}^{4}+{l}^{6}{b}^{4}+{b}^{4
}{r}^{6}+24{l}^{4}{b}^{4}Mr{x}^{4}+24{l}^{4}{b}^{4}Mr{x}^{2}+36{
	l}^{2}{b}^{4}{M}^{2}{r}^{2}{x}^{4}
\]
\be 
\left. -8{l}^{2}{b}^{4}M{r}^{3}{x}^{2}-40
{l}^{2}{b}^{4}M{r}^{3}{x}^{4}+4{l}^{4}{b}^{4}{M}^{2}{x}^{4}-30{l}^{4}{b}^{4}{r}^{2}{x}^{2}-9{l}^{4}{r}^{2}{b}^{4}{x}^{4}-6{l}^{2}{b}^{4}{r}^{4}{x}^{2}+15{l}^{2}{b}^{4}{r}^{4}{x}^{4}+{r}^{2}+{l}^{2} \right\}
\,.\ee 

On the other hand, the following functions appeared in (\ref{Ex})
\be 
h_{10} = -{b}^{8} \left( 3{x}^{4}+6{x}^{2}-1 \right)  \left( 1+3{x}^{2}
\right) ^{2}
\,,\ee 
\[
h_8 = -{b}^{6} \left( 28{b}^{2}{x}^{4}Mr-72{b}^{2}{r}^{2}{x}^{6}+80{x}^{6}{b}^{2}{M}^{2}-66{b}^{2}{r}^{2}{x}^{4}+32{b}^{2}{M}^{2}{x}^{4}+36{b}^{2}Mr{x}^{2}+48{x}^{8}{b}^{2}{M}^{2}\right.
\]
\be 
\left. +36{b}^{2}Mr{x}^{8}-{b}^{2}{r}^{2}-144{b}^{2}{r}^{2}{x}^{2}-72{x}^{6}-60{x}^{4}+4+156
{b}^{2}{x}^{6}Mr+27{b}^{2}{x}^{8}{r}^{2} \right) 
\,,\ee 
\[
h_6 = -2{b}^{4} \left( 2{b}^{2}{r}^{2}+48{b}^{4}M{r}^{3}{x}^{8}+536
{b}^{4}{x}^{6}M{r}^{3}-312{b}^{4}{x}^{6}{M}^{2}{r}^{2}+40{b}^{4}
{x}^{8}{M}^{3}r -3+136{b}^{4}{M}^{3}{x}^{6}r+8{b}^{4}{M}^{4}{x}^{8}\right.
\]
\[
-9
{b}^{4}{r}^{4}{x}^{8}-336{b}^{4}{x}^{6}{r}^{4}-12{b}^{2}{x}^{4}M
r-78{b}^{2}{x}^{6}Mr-38{b}^{2}Mr{x}^{2}-40{b}^{4}M{r}^{3}{x}^{2}
-544{b}^{4}M{r}^{3}{x}^{4}-32{b}^{2}{M}^{2}{x}^{4}-40{x}^{6}{b}^
{2}{M}^{2}
\]
\be 
\left. -18{b}^{2}{r}^{2}{x}^{4}+18{b}^{2}{r}^{2}{x}^{6}+19{b}
^{4}{r}^{4}-72{b}^{4}{r}^{4}{x}^{2}+398{b}^{4}{r}^{4}{x}^{4}+15{x}^{4}+136{b}^{4}{M}^{2}{r}^{2}{x}^{4}+126{b}^{2}{r}^{2}{x}^{2}
\right) 
\,,\ee 
\[
h_4 = -2{b}^{2} \left( 2-12{r}^{5}{b}^{6}M{x}^{2}+31{b}^{6}{r}^{6}-3
{b}^{2}{r}^{2}-170{b}^{4}{x}^{6}M{r}^{3}+88{b}^{4}{x}^{6}{M}^{2}{r
}^{2}-104{b}^{4}{M}^{3}{x}^{6}r-180{b}^{6}M{r}^{5}{x}^{8} \right.
\]
\[
+264{b}
^{6}{M}^{2}{r}^{4}{x}^{8}-144{b}^{6}{M}^{3}{r}^{3}{x}^{8}+72{b}^{6}{M}^{4}{r}^{2}{x}^{8}+102{b}^{4}{x}^{6}{r}^{4}-2{b}^{2}{x}^{4}Mr+
22{b}^{2}Mr{x}^{2}+110{b}^{4}M{r}^{3}{x}^{2}+252{b}^{4}M{r}^{3}{x}^{4}
\]
\[
+16{b}^{2}{M}^{2}{x}^{4}-516{b}^{6}M{r}^{5}{x}^{6}+324{b}^
{6}M{r}^{5}{x}^{4}+51{b}^{6}{r}^{6}{x}^{8}+27{b}^{2}{r}^{2}{x}^{4}
-14{b}^{4}{r}^{4}+34{b}^{4}{r}^{4}{x}^{2}-186{b}^{4}{r}^{4}{x}^{4}-120{b}^{4}{M}^{2}{r}^{2}{x}^{4}
\]
\be 
\left. -336{r}^{4}{b}^{6}{M}^{2}{x}^{4}
-112{b}^{6}{M}^{3}{x}^{6}{r}^{3}+504{b}^{6}{M}^{2}{x}^{6}{r}^{4}+
64{b}^{6}{r}^{6}{x}^{6}+70{b}^{6}{r}^{6}{x}^{4}-88{b}^{6}{r}^{6}
{x}^{2}-60{b}^{2}{r}^{2}{x}^{2} \right) 
\,,\ee 
\[
h_2 = 1-96{b}^{8}{r}^{7}{x}^{8}M+48{b}^{8}M{r}^{7}{x}^{2}-512{b}^{8}{r}^{7}{x}^{4}M+48{b}^{8}{r}^{7}{x}^{6}M-140{r}^{5}{b}^{6}M{x}^{2}-
27{b}^{8}{r}^{8}+36{b}^{6}{r}^{6}-4{b}^{2}{r}^{2}
\]
\[
+4{b}^{2}Mr{x}^{2}+88{b}^{4}M{r}^{3}{x}^{2}+40{b}^{4}M{r}^{3}{x}^{4}-28{b}^{6}M{r}^{5}{x}^{6}+296{b}^{6}M{r}^{5}{x}^{4}-6{b}^{4}{r}^{4}+16{b}
^{4}{r}^{4}{x}^{2}-50{b}^{4}{r}^{4}{x}^{4}-64{b}^{4}{M}^{2}{r}^{2}
{x}^{4}
\]
\[
-208{r}^{4}{b}^{6}{M}^{2}{x}^{4}+144{b}^{6}{M}^{3}{x}^{6}{r}^{3}-64{b}^{6}{M}^{2}{x}^{6}{r}^{4}-144{b}^{8}{M}^{3}{r}^{5}{x}^{8}+208{b}^{8}{r}^{6}{x}^{6}{M}^{2}+192{b}^{8}{r}^{6}{x}^{8}{M}^{2}
+272{b}^{8}{r}^{6}{M}^{2}{x}^{4}
\]
\be 
-144{b}^{8}{r}^{5}{M}^{3}{x}^{6}+
122{b}^{8}{r}^{8}{x}^{4}-24{b}^{8}{r}^{8}{x}^{6}+48{b}^{8}{r}^{8}{x}^{2}+9{b}^{8}{r}^{8}{x}^{8}-28{b}^{6}{r}^{6}{x}^{6}-4{b}^{6}
{r}^{6}{x}^{4}-4{b}^{6}{r}^{6}{x}^{2}-12{b}^{2}{r}^{2}{x}^{2}
\,,\ee 
\[
h_0 = -{r}^{2} \left( 1+{b}^{2}{r}^{2}{x}^{2}-{b}^{2}{r}^{2} \right) ^{2}
\left( 6{b}^{4}{r}^{4}{x}^{2}+3{b}^{4}{r}^
{4}+4{b}^{4}M{r}^{3}{x}^{4}-12{b}^{4}M{r}^{3}{x}^{2}-2{b}^{2}{r}
^{2}{x}^{2}\right. 
\]
\be 
\left.-{b}^{4}{r}^{4}{x}^{4} -2{b}^{2}{r}^{2}+12{b}^{2}Mr{x}^{2}-1 \right) 
\,,\ee 
\[
\Xi_x= -16{b}^{2}{l}^{2}Mr{x}^{2}+12{b}^{2}{l}^{2}{r}^{2}{x}^{2}-4{b}^{2}{l}^{2}{r}^{2}-6{b}^{2}{l}^{4}{x}^{2}+2{b}^{2}{r}^{4}{x}
^{2}-2{b}^{2}{l}^{4}-2{b}^{2}{r}^{4}-2{b}^{4}{r}^{6}{x}^{2}+{b}^
{4}{r}^{6}{x}^{4}
\]
\[
+9{l}^{6}{b}^{4}{x}^{4}+6{l}^{6}{b}^{4}{x}^{2}+7
{l}^{4}{b}^{4}{r}^{2}+7{l}^{2}{b}^{4}{r}^{4}+{l}^{6}{b}^{4}+{b}^{4}{r}^{6}+24{l}^{4}{b}^{4}Mr{x}^{4}+24{l}^{4}{b}^{4}Mr{x}^{2}+36{l}^{2}{b}^{4}{M}^{2}{r}^{2}{x}^{4}
\]
\be 
-8{l}^{2}{b}^{4}M{r}^{3}{x}^{2}-40
{l}^{2}{b}^{4}M{r}^{3}{x}^{4}+4{l}^{4}{b}^{4}{M}^{2}{x}^{4}-30{l}^{4}{b}^{4}{r}^{2}{x}^{2}-9{l}^{4}{r}^{2}{b}^{4}{x}^{4}-6{l}^{2}{b}^{4}{r}^{4}{x}^{2}+15{l}^{2}{b}^{4}{r}^{4}{x}^{4}+{r}^{2}+{l}^{2}\,.
\ee 

Furthermore, these functions belong to (\ref{Br})
\be 
{\tilde f}_{12} = -{b}^{8} \Delta_x  \left( 3{x}^{2}+5
\right)  \left( 1+3{x}^{2} \right) ^{2}
\,,\ee 
\[
{\tilde f}_{10} = -4{b}^{6} \left( 20{b}^{2}{r}^{2}{x}^{2}+81{b}^{2}{x}^{8}{r}^{2} -117{b}^{2}Mr{x}
^{8}-14{x}^{2}+88{b}^{2}{x}^{4}Mr+6{b}^{2}{x}^{6}Mr-54{b}^{2}{r}^{2}{x}^{4}-2{b}^{2}{M}^{2}{x}^{4} \right.
\]
\be 
\left. +5{b}^{2}Mr
-36{b}^{2}{r}^{2}{x}^{6}-2{x}^{6}{b}^{2}{M}^{2}-11{b}^{2}{r}^{2}
+18{x}^{6}+18{x}^{8}{b}^{2}{M}^{2}+18{b}^{2}Mr{x}^{2}-4+2{b}^{2}{M}^{2}{x}^{2} \right) 
\,,\ee 
\[
{\tilde f}_8 = -{b}^{4} \left( 18+32{b}^{4}{M}^{4}{x}^{6}+92{b}^{2}{r}^{2}+128{b}^{4}{x}^{4}{M}^{3}r+256{b}^{4}{x}^{2}{M}^{2}{r}^{2}-1200{b}^{4}{M}^{2}{x}^{8}{r}^{2}+1584{b}^{4}M{r}^{3}{x}^{8} \right.
\]
\[
-544{b}^{4}{x}^{6}M
{r}^{3}+288{b}^{4}{x}^{6}{M}^{2}{r}^{2}-96{b}^{4}{x}^{8}{M}^{3}r+
32{b}^{4}{M}^{3}{x}^{6}r+80{b}^{4}{M}^{4}{x}^{8}-387{b}^{4}{r}^{4}{x}^{8}-84{b}^{4}{x}^{6}{r}^{4}
\]
\[
-256{b}^{2}{x}^{4}Mr+480{b}^{2}
{x}^{6}Mr-160{b}^{2}Mr{x}^{2}-24{b}^{2}{M}^{2}{x}^{2}-224{b}^{4}
M{r}^{3}{x}^{2}-832{b}^{4}M{r}^{3}{x}^{4}+16{b}^{4}M{r}^{3}
\]
\[
-64{b}^{2}Mr-8{x}^{6}{b}^{2}{M}^{2}+276{b}^{2}{r}^{2}{x}^{4}-252{b}^{2}{r}^{2}{x}^{6}-131{b}^{4}{r}^{4}+172{b}^{4}{r}^{4}{x}^{2}+430{b}^{4}{r}^{4}{x}^{4}+28{x}^{2}
\]
\be 
\left. -30{x}^{4}+592{b}^{4}{M}^{2}{r}^{2}{x}^{4}-116{b}^{2}{r}^{2}{x}^{2} \right) 
\,,\ee 
\[
{\tilde f}_6 = -8{b}^{2} \left( 93{b}^{4}{x}^{6}{M}^{2}{r}^{2}-1-94{r}^{5}{b}^{6}M{x}^{2}-4{b}^{6}{r}^{6}-4
{b}^{4}{M}^{4}{x}^{6}-9{b}^{2}{r}^{2}-24{b}^{4}{x}^{4}{M}^{3}r-63
{b}^{4}{x}^{2}{M}^{2}{r}^{2}\right.
\]
\[
-92{b}^{4}{x}^{6}M{r}^{3}+16{b}^{4}{M}^{3}{x}^{6}r+24{b}^{6}{M}^{5}{x}^
{8}r+168{b}^{6}{r}^{3}{M}^{3}{x}^{4}+46{b}^{6}{x}^{2}{M}^{2}{r}^{4}+72{b}^{6}{M}^{4}{x}^{6}{r}^{2}-75{b}^{6}M{r}^{5}{x}^{8}
\]
\[
+192{b}
^{6}{M}^{2}{r}^{4}{x}^{8}-32{b}^{6}{M}^{3}{r}^{3}{x}^{8}-96{b}^{6}
{M}^{4}{r}^{2}{x}^{8}+15{b}^{4}{x}^{6}{r}^{4}-16{b}^{2}{x}^{4}Mr+9
{b}^{2}Mr{x}^{2}+3{b}^{2}{M}^{2}{x}^{2}+56{b}^{4}M{r}^{3}{x}^{2}
\]
\[
+42{b}^{4}M{r}^{3}{x}^{4}-9{b}^{6}M{r}^{5}-6{b}^{4}M{r}^{3}+9{b}^{2}Mr+{b}^{2}{M}^{2}{x}^{4}-282{b}^{6}M{r}^{5}{x}^{6}+460{b}^{6}M{r}^{5}{x}^{4}+3{b}^{2}{r}^{2}{x}^{4}+27{b}^{4}{r}^{4}
\]
\[
-55{b}^{4}{r}^{4}{x}^{2}+13{b}^{4}{r}^{4}{x}^{4}-22{b}^{4}{M}^{2}{r}^{2}{x}^{4}-420{r}^{4}{b}^{6}{M}^{2}{x}^{4}-120{b}^{6}{M}^{3}{x}^{6}{r}^
{3}+174{b}^{6}{M}^{2}{x}^{6}{r}^{4}+104{b}^{6}{r}^{6}{x}^{6}
\]
\be 
\left. -148{b}^{6}{r}^{6}{x}^{4}+48{b}^{6}{r}^{6}{x}^{2}+4{b}^{2}{r}^{2}{x}^{2} \right) 
\,,\ee
\[
{\tilde f}_4 = 416{b}^{8}M{r}^{7}{x}^{2}-640{b}^{8}{r}^{7}{x}^{4}M+864{b}^{8}{r}^{7}{x}^{6}M-544{r}^{5}{b}^{6}M{x}^{2}+33{b}^{8}{r}^{8}-64{b}^{6}{r}^{6}-1-624{b}^{8}{r}^{7}{x}^{8}M
\]
\[
+912{b}^{8}{M}^{4}{r}^{4}{x}
^{8}-1056{b}^{8}{r}^{4}{M}^{4}{x}^{6}-28{b}^{2}{r}^{2}-64{b}^{4}
{x}^{4}{M}^{3}r-272{b}^{4}{x}^{2}{M}^{2}{r}^{2}+960{b}^{6}{r}^{3}{M}^{3}{x}^{4}+456{b}^{6}{x}^{2}{M}^{2}{r}^{4}
\]
\[
+96{b}^{6}{M}^{4}{x}^
{6}{r}^{2}-16{b}^{2}Mr{x}^{2}+8{b}^{2}{M}^{2}{x}^{2}+272{b}^{4}M
{r}^{3}{x}^{2}-256{b}^{4}M{r}^{3}{x}^{4}-96{b}^{6}M{r}^{5}-48{b}
^{4}M{r}^{3}+32{b}^{2}Mr
\]
\[
-16{b}^{8}M{r}^{7}-864{b}^{6}M{r}^{5}{x}
^{6}+1504{b}^{6}M{r}^{5}{x}^{4}+108{b}^{4}{r}^{4}-184{b}^{4}{r}^
{4}{x}^{2}+92{b}^{4}{r}^{4}{x}^{4}+336{b}^{4}{M}^{2}{r}^{2}{x}^{4}
\]
\[
-1088{b}^{6}{M}^{3}{x}^{6}{r}^{3}
+1560{b}^{6}{M}^{2}{x}^{6}{r}^{4}-2208{b}^{8}{M}^{3}{r}^{5}{x}^{8}
-1504{b}^{8}{r}^{6}{x}^{6}{M}^{2}+1680{b}^{8}{r}^{6}{x}^{8}{M}^{2}-240{b}^{8}{r}^{6}{M}^{2}{x}^{4}
\]
\[
+2400{b}^{8}{r}^{5}{M}^{3}{x}^{6}+
518{b}^{8}{r}^{8}{x}^{4}-372{b}^{8}{r}^{8}{x}^{6}-308{b}^{8}{r}^
{8}{x}^{2}+129{b}^{8}{r}^{8}{x}^{8}+224{b}^{6}{r}^{6}{x}^{6}-576
{b}^{6}{r}^{6}{x}^{4}
\]
\be 
+416{b}^{6}{r}^{6}{x}^{2}+12{b}^{2}{r}^{2}{x}
^{2}-1984{r}^{4}{b}^{6}{M}^{2}{x}^{4}
\,,\ee 
\[
{\tilde f}_2 = -4r \left( 1+{b}^{2}{r}^{2}{x}^{2}-{b}^{2}{r}^{2} \right)  \left( 5{r}^{3}{b}^{2}+11{b}^{6}{r}^{7}{x}^{4}-
9{b}^{6}{r}^{7}{x}^{6}-9{b}^{4}{r}^{4}M -
15{b}^{4}{r}^{5}{x}^{4}\right.
\]
\[
-7{b}^{6}{r}^{7}{x}^{2}-3
M{r}^{2}{b}^{2}+35{x}^{6}{b}^{6}M{r}^{6}-15{x}^{4}{b}^{6}M{r}^{6}+30{b}^{6}{M}^{2}{r}^{5}{x}^{2}+40{b}^{4}{M}^{2}{r}^{3}{x}^{2}-32
{b}^{4}{M}^{2}{x}^{4}{r}^{3}
\]
\[
-40{b}^{4}M{x}^{2}{r}^{4}+37{b}^{4}M
{x}^{4}{r}^{4}+11M{r}^{2}{b}^{2}{x}^{2}-6{b}^{2}{M}^{2}r{x}^{2}-15
{b}^{6}{x}^{2}M{r}^{6}-66{b}^{6}{M}^{2}{x}^{6}{r}^{5}+48{b}^{6}{M}^{3}{r}^{4}{x}^{6}
\]
\be 
\left. -48{b}^{6}{M}^{3}{r}^{4}{x}^{4}+36{b}^{6}{M}^{2}{r}^{5}{x}^{4}-{b}^{4}{r}^{5}+5{b}^{6}{r}^{7}-r+M-5{b}^{6}M{r}^{6}-7{b}^{2}{r}^{3}{x}^{2}+20{b}^{4}{r}^{5}{x}^{2} \right) 
\,,\ee
\be 
{\tilde f}_0 = {r}^{4} \left( 1+{b}^{2}{r}^{2}{x}^{2}-{b}^{2}{r}^{2} \right) ^{4}
\,,\ee
\[
\Upsilon_r = -16{b}^{2}{l}^{2}Mr{x}^{2}+12{b}^{2}{l}^{2}{r}^{2}{x}^{2}-4{b}^{2}{l}^{2}{r}^{2}-6{b}^{2}{l}^{4}{x}^{2}+2{b}^{2}{r}^{4}{x}^{2}-2
{b}^{2}{l}^{4}-2{b}^{2}{r}^{4}-2{b}^{4}{r}^{6}{x}^{2}+{b}^{4}{r}^{6}{x}^{4}
\]
\[
+9{l}^{6}{b}^{4}{x}^{4}+6{l}^{6}{b}^{4}{x}^{2}+7{l}^{4}
{b}^{4}{r}^{2}+7{l}^{2}{b}^{4}{r}^{4}+{l}^{6}{b}^{4}+{b}^{4}{r}^{6}+
24{l}^{4}{b}^{4}Mr{x}^{4}+24{l}^{4}{b}^{4}Mr{x}^{2}+36{l}^{2}{b}
^{4}{M}^{2}{r}^{2}{x}^{4}
\]
\be 
-8{l}^{2}{b}^{4}M{r}^{3}{x}^{2}-40{l}^{2}
{b}^{4}M{r}^{3}{x}^{4}+4{l}^{4}{b}^{4}{M}^{2}{x}^{4}-30{l}^{4}{b}^
{4}{r}^{2}{x}^{2}-9{l}^{4}{r}^{2}{b}^{4}{x}^{4}-6{l}^{2}{b}^{4}{r}
^{4}{x}^{2}+15{l}^{2}{b}^{4}{r}^{4}{x}^{4}+{r}^{2}+{l}^{2}\,.
\ee

Finally, these functions appear in (\ref{Bx})
\be 
{\tilde h}_{10} = {b}^{8} \left( 1+3{x}^{2} \right)  \left( 
8M{x}^{4}-9r{x}^{4}-51{x}^{2}r+12M{x}^{2}+5r-9r{x}^{6}+12M{x}^{6} \right)
\,,\ee 
\[
{\tilde h}_8 = {b}^{6} \left( 192{b}^{2}r{x}^{8}{M}^{2}-36M{x}^{6}-16r+117{b}
^{2}{r}^{3}{x}^{8}-36M{x}^{2}+416{b}^{2}{x}^{6}{M}^{2}r+60{b}^{2
}{r}^{3}{x}^{2}-288{b}^{2}{r}^{2}{x}^{8}M\right.
\]
\[
+14{b}^{2}{r}^{3}{x}^{4}+
16{b}^{2}{x}^{8}{M}^{3}-16{b}^{2}{x}^{6}{M}^{3}+96{x}^{2}r-88M
{x}^{4}+96{b}^{2}{M}^{2}r{x}^{4}+540{b}^{2}{r}^{3}{x}^{6}
\]
\be 
\left. -416{b}
^{2}M{r}^{2}{x}^{4}-960{b}^{2}{r}^{2}{x}^{6}M+37{r}^{3}{b}^{2}+240
r{x}^{4} \right) 
\,,\ee 
\[
{\tilde h}_6 = 2{b}^{4} \left( 206{b}^{4}{r}^{5}{x}^{4}-34{r}^{3}{b}^{2}-134{b}^{2}{r}^{3}{x}^{4}-126{b}^{2}{r}^{3}{x}^{6}+8{b}^{2}{x}^{6}{M}^{3}+56{b}^{4}{M}^{2}{x}^{4}{r}^{3}+164{b}^{4}M{x}^{2}{r}^{4}\right.
\]
\[
-380{b}^{4}M{x}^{4}{r}^{4}+10M{r}^{2}{b}^{2}{x}^{2}+25{b}^{4}{r}^{5}-39
r{x}^{4}-42{x}^{2}r+9r+336{b}^{4}{x}^{6}{M}^{3}{r}^{2}-1312{b}^{4}{x}^{6}{M}^{2}{r}^{3}+24{b}^{4}{M}^{4}r{x}^{8}
\]
\[
-408{b}^{4}{x}
^{8}{M}^{2}{r}^{3}+300{b}^{4}{x}^{8}M{r}^{4}+176{b}^{4}{x}^{8}{M}^
{3}{r}^{2}+1388{b}^{4}{r}^{4}{x}^{6}M-104{b}^{2}{M}^{2}r{x}^{4}+
412{b}^{2}M{r}^{2}{x}^{4}-168{b}^{2}{x}^{6}{M}^{2}r
\]
\be 
\left. +234{b}^{2}{r}^{2}{x}^{6}M+22M{x}^{4}-63{b}^{4}{r}^{5}{x}^{8}-444{b}^{4}{r}^{5}{x}^{6}+6{b}^{2}{r}^{3}{x}^{2}-108{b}^{4}{r}^{5}{x}^{2}+18M{x}
^{2} \right) 
\,,\ee
\[
{\tilde h}_4 = 2{b}^{2} \left( 21{r}^{3}{b}^{2}-98{b}^{4}{r}^{5}{x}^{4}+214{b}^{6}{r}^{7}{x}^{4}-44{b}^{6}{r}^{7}{x}^{6}-124{b}^{6}{r}^{7}{x}^
{2}+61{b}^{2}{r}^{3}{x}^{4}-224{x}^{6}{b}^{6}M{r}^{6} \right.
\]
\[
-272{x}^{4}
{b}^{6}M{r}^{6}+136{b}^{4}{M}^{2}{x}^{4}{r}^{3}-110{b}^{4}M{x}^{2}
{r}^{4}+76{b}^{4}M{x}^{4}{r}^{4}-12M{r}^{2}{b}^{2}{x}^{2}+128{b}
^{6}{x}^{2}M{r}^{6}+592{b}^{6}{M}^{2}{x}^{6}{r}^{5}
\]
\[
-360{b}^{6}{M}^
{3}{r}^{4}{x}^{6}+48{b}^{6}{M}^{2}{r}^{5}{x}^{4}+69{b}^{6}{r}^{7}{x}^{8}-38{b}^{4}{r}^{5}+12{x}^{2}r+216{b}^{6}{M}^{4}{r}^{3}{x}^{8}+672{b}^{6}{x}^{8}{M}^{2}{r}^{5}-336{b}^{6}{x}^{8}M{r}^{6}
\]
\[
-600
{b}^{6}{x}^{8}{M}^{3}{r}^{4}+13{b}^{6}{r}^{7}-4r-216{b}^{4}{x}^{6}{M}^{3}{r}^{2}+504{b}^{4}{x}^{6}{M}^{2}{r}^{3}-462{b}^{4}{r}^{4}
{x}^{6}M+56{b}^{2}{M}^{2}r{x}^{4}
\]
\be 
\left. -148{b}^{2}M{r}^{2}{x}^{4}+126{b}^{4}{r}^{5}{x}^{6}-18{b}^{2}{r}^{3}{x}^{2}+106{b}^{4}{r}^{5}{x}^
{2}-6M{x}^{2} \right) 
\,,\ee 
\[
{\tilde h}_2 = r \left( 1+{b}^{2}{r}^{2}{x}^{2}-{b}^{2}{r}^{2} \right)  \left( 25{b}^{6}{r}^{6}{x}^{6}-11{b}^{6}{r}^{6}{x}^{4}+59{b}^{6}{r}^{6}{x}^{2}-9{b}^{6}{r}^{6}-60{b}^{6}M{r}^{5}{x}^{6} -56{b}^{6}M{r}^{5}{x}^
{4}\right.
\]
\[
-44{r}^{5}{b}^{6}M{x}^{2}+48{b}^{6}{M}^{2}{x}^{6}{r}^{4}+48{r}^{4}{b}^{6}{M}^{2}{x}^{4}+35{b}^{4}{r}^{4}{x}^{4}-38{b}^{4}{r}^{4}{x}^{2}+19{b}^{4}{r}^{4}-24{b}^{4}M{r}^{3}{x}^{4}
\]
\be 
\left. +40{b}^{4}M{r}
^{3}{x}^{2}-48{b}^{4}{M}^{2}{r}^{2}{x}^{4}+11{b}^{2}{r}^{2}{x}^{2}
-11{b}^{2}{r}^{2}+4{b}^{2}Mr{x}^{2}+1 \right) 
\,,\ee 
\be 
{\tilde h}_0 = {r}^{3} \left( 1+{b}^{2}{r}^{2}{x}^{2}-{b}^{2}{r}^{2} \right) ^{4}
\,,\ee
\[
\Upsilon_x = -16{b}^{2}{l}^{2}Mr{x}^{2}+12{b}^{2}{l}^{2}{r}^{2}{x}^{2}-4{b}^{2}{l}^{2}{r}^{2}-6{b}^{2}{l}^{4}{x}^{2}+2{b}^{2}{r}^{4}{x}
^{2}-2{b}^{2}{l}^{4}-2{b}^{2}{r}^{4}-2{b}^{4}{r}^{6}{x}^{2}+{b}^
{4}{r}^{6}{x}^{4}+9{l}^{6}{b}^{4}{x}^{4}
\]
\[
+6{l}^{6}{b}^{4}{x}^{2}+7
{l}^{4}{b}^{4}{r}^{2}+7{l}^{2}{b}^{4}{r}^{4}+{l}^{6}{b}^{4}+{b}^{4}{r}^{6}+24{l}^{4}{b}^{4}Mr{x}^{4}+24{l}^{4}{b}^{4}Mr{x}^{2}+36{l}^{2}{b}^{4}{M}^{2}{r}^{2}{x}^{4}-8{l}^{2}{b}^{4}M{r}^{3}{x}^{2}
\]
\be 
-40
{l}^{2}{b}^{4}M{r}^{3}{x}^{4}+4{l}^{4}{b}^{4}{M}^{2}{x}^{4}-30{l}^{4}{b}^{4}{r}^{2}{x}^{2}-9{l}^{4}{r}^{2}{b}^{4}{x}^{4}-6{l}^{2}{b}^{4}{r}^{4}{x}^{2}+15{l}^{2}{b}^{4}{r}^{4}{x}^{4}+{r}^{2}+{l}^{2}
\,.\ee


\begin{thebibliography}{99}
	
\bibitem{Stephani:2003tm} 
H.~Stephani, D.~Kramer, M.~A.~H.~MacCallum, C.~Hoenselaers and E.~Herlt,
``Exact solutions of Einstein's field equations,''
Cambridge University Press, 2004.

\bibitem{IslamBook}
Islam, J. N., 
``Rotating fields in general relativity,'' 
Cambridge University Press (1985).

\bibitem{Wald:1984rg.book} 
R.~M.~Wald,
`General Relativity,''
Chicago, Usa: Univ. Pr. ( 1984) 491p.

\bibitem{Misner:1974qy} 
C.~W.~Misner, K.~S.~Thorne and J.~A.~Wheeler,
``Gravitation,''
San Francisco 1973, 1279p.

\bibitem{Griffiths:2009dfa}
J.~B.~Griffiths and J.~Podolsky,
Cambridge University Press, 2009.

\bibitem{Wald:1974np} 
R.~M.~Wald,
Phys.\ Rev.\ D {\bf 10}, 1680 (1974).

\bibitem{Ernst}
F. J. Ernst,
J. Math. Phys. 17, 54 (1976).

\bibitem{Harrison}
B. K. Harrison, 
J.Math.Phys.,9,1744, (1968).

\bibitem{Aliev:1989wx} 
A.~N.~Aliev and D.~V.~Galtsov,
Sov.\ Phys.\ Usp.\  {\bf 32}, 75 (1989).

\bibitem{Kolos:2015iva} 
M.~Kološ, Z.~Stuchlik and A.~Tursunov,
Class.\ Quant.\ Grav.\  {\bf 32}, no. 16, 165009 (2015).

\bibitem{Tursunov:2014loa} 
A.~Tursunov, M.~Kolos, Z.~Stuchlik and B.~Ahmedov,
Phys.\ Rev.\ D {\bf 90}, no. 8, 085009 (2014).


\bibitem{Aliev:1989wz} 
A.~N.~Aliev and D.~V.~Galtsov,
Astrophys.\ Space Sci.\  {\bf 155}, 181 (1989).

\bibitem{Aliev:1980et} 
A.~N.~Aliev, D.~V.~Galtsov and A.~A.~Sokolov,
Sov.\ Phys.\ J.\  {\bf 23}, 179 (1980).

\bibitem{Aliev:1986wu} 
A.~N.~Aliev, D.~V.~Galtsov and V.~I.~Petrukhov,
Astrophys.\ Space Sci.\  {\bf 124}, 137 (1986).


\bibitem{Aliev:1988wy} 
A.~N.~Aliev and D.~V.~Galtsov,
Sov.\ Phys.\ JETP {\bf 67}, 1525 (1988).


\bibitem{Aliev:1989sw} 
A.~N.~Aliev and D.~V.~Galtsov,
Sov.\ Phys.\ J.\  {\bf 32}, 790 (1989).

\bibitem{Galtsov:1978ag} 
D.~V.~Galtsov and V.~I.~Petukhov,
Zh.\ Eksp.\ Teor.\ Fiz.\  {\bf 74}, 801 (1978).

\bibitem{Hiscock:1980zf} 
W.~A.~Hiscock,
J.\ Math.\ Phys.\  {\bf 22}, 1828 (1981).

\bibitem{Orekhov:2016bpc} 
K.~Orekhov,
J.\ Geom.\ Phys.\  {\bf 104}, 242 (2016).

\bibitem{Gibbons:2013yq} 
G.~W.~Gibbons, A.~H.~Mujtaba and C.~N.~Pope,
Class.\ Quant.\ Grav.\  {\bf 30}, no. 12, 125008 (2013).

\bibitem{Gibbons:2013dna} 
G.~W.~Gibbons, Y.~Pang and C.~N.~Pope,
Phys.\ Rev.\ D {\bf 89}, no. 4, 044029 (2014).

\bibitem{Rogatko:2016knj} 
M.~Rogatko,
Phys.\ Rev.\ D {\bf 93}, no. 4, 044008 (2016).


\bibitem{Bicak:2015lxa} 
J.~Bicak and F.~Hejda,
Phys.\ Rev.\ D {\bf 92}, no. 10, 104006 (2015).

\bibitem{Brito:2014nja} 
R.~Brito, V.~Cardoso and P.~Pani,
Phys.\ Rev.\ D {\bf 89}, no. 10, 104045 (2014).


\bibitem{Siahaan:2015xia} 
H.~M.~Siahaan,
Class.\ Quant.\ Grav.\  {\bf 33}, no. 15, 155013 (2016).


\bibitem{Astorino:2015lca} 
M.~Astorino,
JHEP {\bf 1510}, 016 (2015).

\bibitem{Astorino:2015naa} 
M.~Astorino,
Phys.\ Lett.\ B {\bf 751}, 96 (2015).

\bibitem{Astorino:2016hls} 
M.~Astorino, G.~Compère, R.~Oliveri and N.~Vandevoorde,
Phys.\ Rev.\ D {\bf 94}, no. 2, 024019 (2016).

\bibitem{Chakraborty:2017nfu} 
C.~Chakraborty and S.~Bhattacharyya,
Phys.\ Rev.\ D {\bf 98}, no. 4, 043021 (2018).

\bibitem{Pradhan:2014zia} 
P.~Pradhan,
Class.\ Quant.\ Grav.\  {\bf 32}, no. 16, 165001 (2015).

\bibitem{Sakti:2017pmt}
M.~F.~A.~R.~Sakti, A.~Suroso and F.~P.~Zen,
Int.\ J.\ Mod.\ Phys.\ D {\bf 27} (2018) no.12,  1850109.

\bibitem{Sakti:2019krw}
M.~F.~A.~R.~Sakti, A.~Suroso and F.~P.~Zen,
arXiv:1901.09163 [gr-qc].

\bibitem{Aliev:2008wv} 
A.~N.~Aliev, H.~Cebeci and T.~Dereli,
Phys.\ Rev.\ D {\bf 77}, 124022 (2008).

\bibitem{Duztas:2017lxk} 
K.~D\"{u}zta\c{s},
Class.\ Quant.\ Grav.\  {\bf 35}, no. 4, 045008 (2018).

\bibitem{Cebeci:2015fie} 
H.~Cebeci, N.~\''{O}zdemir and S.~\c{S}entorun,
Phys.\ Rev.\ D {\bf 93}, no. 10, 104031 (2016).

\bibitem{Zakria:2015eua} 
A.~Zakria and M.~Jamil,
JHEP {\bf 1505}, 147 (2015).

\bibitem{Mukherjee:2018dmm}
S.~Mukherjee, S.~Chakraborty and N.~Dadhich,
Eur.\ Phys.\ J.\ C {\bf 79} (2019) no.2,  161.

\bibitem{Abdujabbarov:2011uv}
A.~A.~Abdujabbarov, B.~J.~Ahmedov, S.~R.~Shaymatov and A.~S.~Rakhmatov,
Astrophys.\ Space Sci.\  {\bf 334} (2011) 237.

\bibitem{Abdujabbarov:2012jj}
A.~A.~Abdujabbarov, A.~A.~Tursunov, B.~J.~Ahmedov and A.~Kuvatov,
Astrophys.\ Space Sci.\  {\bf 343} (2013) 173.

\bibitem{Ahmedov:2011wb}
B.~J.~Ahmedov, A.~V.~Khugaev and A.~A.~Abdujabbarov,
Astrophys.\ Space Sci.\  {\bf 337} (2012) 19.

\bibitem{Abdujabbarov:2008mz}
A.~A.~Abdujabbarov, B.~J.~Ahmedov and V.~G.~Kagramanova,
Gen.\ Rel.\ Grav.\  {\bf 40} (2008) 2515.

\bibitem{Abdujabbarov:2012bn}
A.~Abdujabbarov, F.~Atamurotov, Y.~Kucukakca, B.~Ahmedov and U.~Camci,
Astrophys.\ Space Sci.\  {\bf 344} (2013) 429.

\bibitem{Jefremov:2016dpi}
P.~Jefremov and V.~Perlick,
Class.\ Quant.\ Grav.\  {\bf 33} (2016) no.24,  245014.

\bibitem{Siahaan:2019kbw}
H.~M.~Siahaan,
Eur. Phys. J. C \textbf{80} (2020) no.10, 1000.

\bibitem{Siahaan:2020bga}
H.~M.~Siahaan,
Phys. Rev. D \textbf{102} (2020) no.6, 064022.

\bibitem{Ciambelli:2020qny}
L.~Ciambelli, C.~Corral, J.~Figueroa, G.~Giribet and R.~Olea,
Phys. Rev. D \textbf{103} (2021) no.2, 024052

\bibitem{Cisterna:2021xxq}
A.~Cisterna, A.~Neira-Gallegos, J.~Oliva and S.~C.~Rebolledo-Caceres,
[arXiv:2101.03628 [gr-qc]].

\bibitem{Frolov:2017bdq}
V.~P.~Frolov, P.~Krtous and D.~Kubiznak,
Phys. Lett. B \textbf{771} (2017), 254-256.

\bibitem{Wild:1980zz}
W.~J.~Wild and R.~M.~Kerns,
Phys. Rev. D \textbf{21} (1980), 332-335.

\bibitem{ErnstI:1967wx} 
F.~J.~Ernst,
Phys.\ Rev.\  {\bf 167}, 1175 (1968).

\bibitem{ErnstII:1967by} 
F.~J.~Ernst,
Phys.\ Rev.\  {\bf 168}, 1415 (1968).

\bibitem{Melvin:1963qx} 
M.~A.~Melvin,
Phys.\ Lett.\  {\bf 8}, 65 (1964).

\bibitem{Booth:2015nwa} 
I.~Booth, M.~Hunt, A.~Palomo-Lozano and H.~K.~Kunduri,
Class.\ Quant.\ Grav.\  {\bf 32}, no. 23, 235025 (2015).

\bibitem{Xulu:1999gf}
S.~S.~Xulu,
Int. J. Mod. Phys. A \textbf{15} (2000), 4849-4856.

\bibitem{Xulu:1999tx}
S.~S.~Xulu,
Int. J. Mod. Phys. A \textbf{15} (2000), 2979-2986.

\bibitem{Thorne:1965}
K.~Thorne, 
Phys. Rev. {\bf 139} (1965) B244.
	
\end{thebibliography}
\end{document}